\documentclass[lettersize,journal]{IEEEtran}
\usepackage{amsmath,amsfonts, amsthm, amssymb, pifont}
\usepackage[noend]{algpseudocode}
\usepackage{algorithm}
\usepackage{multirow, multicol}
\usepackage{booktabs}
\usepackage{array}
\usepackage[caption=false,font=normalsize,labelfont=sf,textfont=sf]{subfig}
\usepackage{textcomp}
\usepackage{stfloats}
\usepackage{url}
\usepackage{verbatim}
\usepackage{graphicx}
\usepackage{cite}
\usepackage{caption}
\usepackage{xparse}
\usepackage{xcolor}
\usepackage{color, colortbl}

\definecolor{Gray}{gray}{0.9}

\newcommand*\Let[2]{\State #1 $\gets$ #2}
\algrenewcommand\algorithmicrequire{\textbf{Precondition:}}
\algrenewcommand\algorithmicensure{\textbf{Postcondition:}}

\newcommand{\eref}[1]{(\ref{#1})}
\newcommand{\sref}[1]{Section~\ref{#1}}
\newcommand{\fref}[1]{Fig.~\ref{#1}}
\newcommand{\tref}[1]{Table~\ref{#1}}
\newcommand{\aref}[1]{Algorithm~\ref{#1}}

\begin{document}

\title{Meta-AF: Meta-Learning for Adaptive Filters}

\author{Jonah Casebeer,~\IEEEmembership{Student Member, IEEE}, Nicholas J. Bryan,~\IEEEmembership{Member, IEEE}, Paris Smaragdis,~\IEEEmembership{Fellow, IEEE}%
\thanks{J. Casebeer is with the Department of Computer Science, University of Illinois at Urbana-Champaign, Urbana, IL 61801 USA~(e-mail: jonahmc2@illinois.edu). Work performed in part while interning at Adobe Research.\\
N. J. Bryan was the lead advisor for this work and is with Adobe Research, San Francisco, CA, 94103 USA~(e-mail: njb@ieee.org)\\
P. Smaragdis is with the Department of Computer Science and Department, University of Illinois at Urbana-Champaign, Urbana, IL 61801 USA~(e-mail: paris@illinois.edu)
Partially funded by NIFA award \#2020-67021-32799}
\thanks{Manuscript received Month Day, Year; revised Month Day, Year.}}

\markboth{}%
{Shell \MakeLowercase{\textit{et al.}}: A Sample Article Using IEEEtran.cls for IEEE Journals}


\maketitle

\begin{abstract}
Adaptive filtering algorithms are pervasive throughout signal processing and have had a material impact on a wide variety of domains including audio processing, telecommunications, biomedical sensing, astrophysics and cosmology, seismology, and many more. Adaptive filters typically operate via specialized online, iterative optimization methods such as least-mean squares or recursive least squares and aim to process signals in unknown or nonstationary environments. Such algorithms, however, can be slow and laborious to develop, require domain expertise to create, and necessitate mathematical insight for improvement. In this work, we seek to improve upon hand-derived adaptive filter algorithms and present a comprehensive framework for learning online, adaptive signal processing algorithms or update rules directly from data. To do so, we frame the development of adaptive filters as a meta-learning problem in the context of deep learning and use a form of self-supervision to learn online iterative update rules for adaptive filters. To demonstrate our approach, we focus on audio applications and systematically develop meta-learned adaptive filters for five canonical audio problems including system identification, acoustic echo cancellation, blind equalization, multi-channel dereverberation, and beamforming. We compare our approach against common baselines and/or recent state-of-the-art methods. We show we can learn high-performing adaptive filters that operate in real-time and, in most cases, significantly outperform each method we compare against -- all using a single general-purpose configuration of our approach.
\end{abstract}


\section{Introduction}
\IEEEPARstart{A}{daptive} signal processing and adaptive filter theory are cornerstones of modern signal processing and have had a deep and significant impact on modern society. Applications of adaptive filters~(AF) include audio processing, telecommunications, biomedical sensing, astrophysics and cosmology, seismology and more. Audio applications, in particular, are of exceptional importance and find utility for many problems such as single- and multi-channel system identification, echo cancellation, prediction, dereverberation, beamforming, noise cancellation, and beyond. AFs typically operate via online iterative optimization methods, such as least mean square filtering~(LMS), normalized LMS~(NLMS), recursive least-squares~(RLS), or Kalman filtering~(KF), to solve streaming or online optimization problems~\cite{Widrow1985, mathews1991adaptive, soo1990multidelay, haykin2008adaptive, apolinario2009qrd, rabiner2016theory} and process signals in unknown and/or nonstationary environments. 

AF tasks are often grouped into one of four core categories: system identification, inverse modeling, prediction, and interference cancellation~\cite{haykin2008adaptive}. 
Each of these categories has numerous AF applications, and advances in one category or application can often be applied to many others. In the audio domain, acoustic echo cancellation~(AEC) can be formulated as single- or multi-channel system identification and has been studied extensively~\cite{schraudolph1999local, gay1998efficient, benesty2001advances, hansler2005acoustic, enzner2006frequency, valin2007adjusting, malik2008model, malik2010online, kuech2014state, yang2017frequency, valero2017multi, haubner2021noise}. Equalization can be formulated as an inverse modeling problem, has been explored in single- and multi-channel formats, and is particularly useful for sound zone reproduction and active noise control~\cite{nelson1992adaptive, nelson1995inverse, bouchard2000multichannel, george2013advances, lu2021survey, lu2021active}. Dereverberation can be formulated as a prediction problem and has been studied considerably in this context~\cite{yoshioka2009adaptive, nakatani2010speech, yoshioka2012generalization, li2017acoustic, caroselli2017adaptive, drude2018nara, wung2020robust}. Finally, multi-microphone enhancement or beamforming can be formulated as an informed interference cancellation task and also has a breadth of associated algorithms~\cite{griffiths1982alternative, hoshuyama1999robust, chen2008minimum, souden2009optimal, huang2011multi, markovich2012sparse, gannot2017consolidated}.

When we consider AFs in the context of deep neural networks~(DNNs), we note two main observations. First, AFs continue to be used extensively, but mostly via hybrid approaches that combine neural networks with standard AF algorithms. Second, the underlying AF algorithms and tools for the development of new AFs have had little change in several decades. Hybrid approaches, however, have proven very successful. For example, in AEC, neural networks can be trained for residual echo suppression, noise suppression, reference estimation, and more~\cite{birkett1995acoustic, rabaa1998acoustic, malek2016hammerstein, zhang2017recursive, halimeh2019neural, zhang2019deep, ma2020acoustic, ivry2021nonlinear, valin2021low}. In a similar vein, neural networks paired with AFs for active noise control tasks have shown impressive results~\cite{zhou2005analysis, krukowicz2010active, zhang2020deep, zhang2021deep}. For dereverberation, DNNs have proven useful for both online and offline approaches by directly estimating statistics of the dereverberated speech~\cite{kinoshita2017neural, heymann2018frame, drude2018integrating, wang2021convolutive, wang2021prediction}. This pattern has repeated itself for beamforming applications, where DNNs have led to many performance improvements~\cite{erdogan2016improved, heymann2016neural, qian2018deep, zhang2021adl}. In many of these works, DNNs are trained to predict ratio-masks, or otherwise directly enhance/separate the desired signal. In essence, they act as a distinct module within a larger pipeline that also uses AFs.

In contrast, there are a small number of works that more tightly couple neural networks and AFs and use DNNs for optimal control of AFs. Recently, it was shown that DNNs can estimate statistics to control step-sizes~\cite{haubner2021end, zhang2022deep} or estimate entire updates~\cite{casebeer2021auto} for a single-channel AEC. Similarly, past work has used DNNs to predict updates for the internal statistics of multi-channel beamformers~\cite{casebeer2021nice} and to learn source-models for multi-channel source separation~\cite{scheibler2021surrogate}. These works differ from hybrid approaches in that they leverage neural networks to update or control AFs directly and thus focus on improving the performance of AFs themselves. Such improvement can be leveraged in isolation or, in theory, together with complementary hybrid approaches.

\IEEEpubidadjcol 
More broadly, the idea of controlling AFs via neural networks is related to several disciplines of signal processing and machine learning including optimal control, optimization, automatic machine learning, reinforcement learning and meta-learning. Relevant works within these areas include automatic selection of step sizes~\cite{sutton1992adapting, schraudolph1999local, mahmood2012tuning}, the direct control of model weights~\cite{schmidhuber1992learning, bello2017neural, ha2016hypernetworks}, rapid fine-tuning~\cite{finn2017model}, and meta-learning optimization rules~\cite{andrychowicz2016learning}. Out of these works, learning optimization rules or \emph{learned optimizers} is of critical relevance~\cite{wichrowska2017learned, metz2019understanding, chen2020training} and presents the idea of using one neural network as a function that optimizes another function. Such works, however, focus on creating learned optimizers for training neural networks in an offline setting, where the latter network is the final product, and the learned optimizer is otherwise discarded (or otherwise used to train additional networks). Moreover, this work has had little application to AFs, except for our own work~\cite{casebeer2021auto}, which we extend here.

In this work, we formulate the development of AF algorithms as a meta-learning problem. We learn adaptive filter update rules directly from data using self-supervision and call our approach meta adaptive filtering~(Meta-AF). Using our method, we no longer need humans to hand-derive update equations, do not need any supervised labeled data for learning, and do not need exhaustive tuning.
To showcase our approach, we systematically develop learned AFs for exemplary applications of each of the four canonical AF architectures. Then, for each AF task, we compare our work to a suite of baselines and/or state-of-the-art approaches for the problems of system identification, acoustic echo cancellation, equalization, multi-channel dereverberation, and multi-channel interference cancellation (beamforming). For all tasks, we use identical hyperparameters, significantly reducing engineering and design time. We evaluate performance using signal-to-noise ratio~(SNR)-like signal metrics and perceptual- or task-specific metrics as well as specific qualitative comparisons. Our results show we can learn high-performing AF algorithms that operate in real-time and, in most cases, outperform all methods we compare against.

The contributions of our work are as follows:
1) we present the first general-purpose method of meta-learning AF algorithms~(update rules) directly from data via self-supervision~(no labels required)
2) we apply our approach to all canonical AF architecture categories including system identification, inverse modeling, prediction, and~(informed) interference cancellation and
3) we show how a single hyperparameter configuration of our approach, trained with different datasets and losses, can outperform all common AF baselines and/or several past state-of-the-art methods we compare against according to one or more evaluation metrics and are suitable for real-time operation on commodity hardware.
Compared to our previous work on single-channel single-talk AEC~\cite{casebeer2021auto}, we present several new improvements including 1) an improved loss, 2) additional inputs to our learned optimizer and 3) an updated development for multi-block, multi-channel AFs with coupled updates and 4) extensive experimentation on four new applications. We release demos for each task and open source all code\footnote{For demos and code, please see \url{https://jmcasebeer.github.io/projects/metaaf} and \url{https://github.com/adobe-research/MetaAF}, respectively.} including baselines.
\def\time{{\mathrm{t}}}
\def\freq{\mathrm{k}}
\def\mic{{\mathrm{m}}}
\def\buffer{\mathrm{b}}
\def\frame{\tau}

\def\F{\mathbf{F}}
\def\I{\mathbf{I}}
\def\0{\mathbf{0}}
\def\1{\mathbf{1}}

\def\u{\mathbf{u}}
\def\U{\mathbf{U}}
\def\d{\mathbf{d}}
\def\D{\mathbf{D}}
\def\y{\mathbf{y}}
\def\e{\mathbf{e}}
\def\s{\mathbf{s}}
\def\w{\mathbf{w}}
\def\vcat{\mathrm{cat}}

\def\H{\mathsf{H}}
\def\T{\top}
\def\diag{\operatorname{diag}}

\def\L{\mathcal{L}}
\def\grad{{\boldsymbol{\nabla}}}
\def\bdelta{{\boldsymbol{\Delta}}}
\def\btheta{{\boldsymbol{\theta}}}
\def\bphi{{\boldsymbol{\phi}}}
\def\bxi{{\boldsymbol{\xi}}}


\section{Background}

\vspace{-.4cm}

\subsection{Notation}
\label{sec:notation}
\begin{table}
\begin{center}
\caption{Symbols and operators.}
\begin{tabular}{ |c|c| } 
\hline
Symbols & Description \\
\hline
$x \in \mathbb{R}$ & A real-valued scalar \\
$x \in \mathbb{C}$ & A complex-valued scalar \\
$\underline{x} \in \mathbb{R}$ & A real-valued time-domain scalar \\
$\underline{\mathbf{x}} \in \mathbb{R}^N$ & A time-domain $N$-dimensional column vector \\
$\mathbf{x} \in \mathbb{C}^N$ & A complex-valued $N$-dimensional column vector \\
$\mathbf{X} \in \mathbb{C}^{M \times N}$ & A complex-valued $M \times N$ matrix \\
$\mathbf{x}[\frame]$ & A time-varying column vector \\
$\mathbf{X}[\frame]$ & A time-varying matrix \\
$\w[\frame]$ & FD AF linear-filter \\ 
$\u[\frame]$ & FD AF input  \\ 
$\d[\frame]$ &  FD AF target or desired response \\
$\y[\frame]$ &  FD AF estimated response \\
$\s[\frame]$ &  FD AF true desired signal \\
$\I_N$ & An $N \times N$ identity matrix \\
$\0_{N \times R}$ & $N \times R$ matrix of zeros \\
$\1_{N \times R}$ & $N \times R$ matrix of ones \\
$\F_N$ & $N$-point discrete Fourier transform~(DFT) matrix \\
\hline
Operators & Description \\
\hline
$\ast$ & Convolution \\
$\cdot^\T$ & Transpose \\
$\cdot^{*}$ & Complex conjugate \\
$\cdot^\H$ & Hermitian transpose \\ 
$\diag(\cdot)$ & Vector to a diagonal matrix \\
$\vcat(\cdots)$ & Column vector concatenation (vertical stack) \\ 
$E$ & Expected value \\
$\frac{(\cdot)}{(\cdot)} $ &  Element-wise division \\
$|| \cdot || $ & Euclidean norm \\
$| \cdot | $ & Element-wise magnitude of complex value \\
$\angle$ & Phase of a complex-value \\
$\ln$ & Natural logarithm \\
${}^{-1}$ & Scalar or matrix inverse \\
\hline
\end{tabular}
\label{tab:notation}
\end{center}
\vspace{-.15cm}
\end{table}

We provide an overview of the symbols and operators we use in~\tref{tab:notation}. We denote scalars via lower-case symbols, column vectors via bold, lower-case symbols, and matrices via bold upper-case symbols. We use bracket indexing $[\frame]$ to denote time-varying signals and an underline to denote the time-domain counterpart of a frequency-domain (FD) symbol. We index FD rows via the subscript $\freq$, columns via the subscript $\mic$, and elements via the subscripts $\freq\mic$.

\subsection{Overlap-Save \& Overlap-Add Filtering}
We perform short-time~(STFT) Fourier filtering using either overlap-save~(OLS) or overlap-add~(OLA) convolution. The OLS method uses block processing by splitting the input signal into overlapping windows and recombines complete non-overlapping components. We use $\underline{u}_{\mic}[\mathrm{t}] \in \mathbb{R}$ to represent the time-domain sample recorded by microphone $\mic$ at discrete time $\mathrm{t}$. We collect $N$ such samples from microphone $\mic$ to form the time-domain frame $\underline{\u}_{\mic}[\frame] = [\underline{u}_{\mic}[\frame R - N + 1], \cdots, \underline{u}_{\mic}[\frame R]] \in \mathbb{R}^{N}$, where $\frame$ is the frame index, $N$ is the window length in samples, $R$ is the hop size in samples, and $O=N-R$ is the overlap between frames in samples. Finally, we collect samples from all $M$ microphones to form a multi-channel time-domain signal, $\underline{\U}[\frame] = [\underline{\u}_1[\frame], \cdots, \underline{\u}_M[\frame]] \in \mathbb{R}^{N \times M}$. We compute the corresponding FD representation via $\U[\frame] = \F_N \underline{\U}[\frame] \in \mathbb{C}^{K \times M}$, where $K$ is the number of frequency bins, set to $K=N$ for this work. We select the $\mic^{\text{th}}$ channel from a multi-channel FD representation using $\u_{\mic}[\frame] \in \mathbb{C}^{K}$. We define the FD filter $\w_{\mic}[\frame] \in \mathbb{C}^{K}$ and the frequency and time output for the $\mic^{\text{th}}$ channel as
\begin{eqnarray}
    \y_{\mic}[\frame] &=& \diag(\u_{\mic}[\frame]) \mathbf{Z}_w \w^*_{\mic}[\frame] \in\mathbb{C}^{K}\label{eq:ols_output} \\
        \underline{\y}_{\mic}[\frame] &=& \mathbf{Z}_y \y_{\mic}[\frame] \in \mathbb{R}^R, \label{eq:ols_output_time}
\end{eqnarray}
where $\mathbf{Z}_w = \F_K \mathbf{T}_{K/2}^\T \mathbf{T}_{K/2} \F_K^{-1}\in\mathbb{C}^{K \times K}$ and $\mathbf{Z}_y = \bar{\mathbf{T}}_{R} \F_K^{-1} \in \mathbb{C}^{R \times K}$ are anti-aliasing matrices, $\mathbf{T}_{K/2} = [\I_{K/2}, \0_{K/2 \times K/2}] \in \mathbb{R}^{R \times K}$ trims the last $K/2$ samples from a vector and $\bar{\mathbf{T}}_{R} = [\0_{R \times O}, \I_{R}] \in \mathbb{R}^{R \times K}$ trims the first $O$ samples.

The counterpart to OLS is OLA filtering, which computes the frequency output, time output, and buffer $\underline{\mathbf{b}}_{\mic}[\frame]$ as
\begin{align}
    \y_{\mic}[\frame] &= \diag(\u_{\mic}[\frame]) \w^*_{\mic}[\frame] \in \mathbb{C}^{K} \label{eq:ola_freq}\\
    \underline{\y}_{\mic}[\frame] &= \mathbf{T}_R \F_K^{-1} \y_{\mic}[\frame] + \bar{\mathbf{T}}_R \underline{\mathbf{b}}_{\mic}[\frame - 1] \in \mathbb{R}^{R} \label{eq:ola_output}\\
    \underline{\mathbf{b}}_{\mic}[\frame] &= \F_K^{-1} \y_{\mic}[\frame] + 
                                 \mathbf{T}_R^\T \bar{\mathbf{T}}_R \underline{\mathbf{b}}_{\mic}[\frame - 1] \in \mathbb{R}^{K}. \label{eq:ola_buffer}
\end{align}
Here, $\mathbf{T}_{R} = [\I_{R}, \0_{R \times O}] \in \mathbb{R}^{R \times K}$.
Typically, the forward and inverse DFTs are combined with analysis and synthesis windows and optionally zero-padded. We use Hann windows~\cite{oppenheim1975digital}. 
For multi-channel multi-block input, single-channel output FD filters, the OLS/OLA equations above are applied per channel, and anti-aliasing is applied per block. The per frequency (anti-aliased)~filter is $\w_{\freq}[\frame] \in \mathbb{C}^{BM}$ with $B$ buffered frames and $M$ channels all stacked. The filter input is similarly stacked and is $\u_\freq[\frame] \in \mathbb{C}^{BM}$, thus requiring~\eqref{eq:ols_output} and~\eqref{eq:ola_freq} to be modified.

\subsection{Adaptive Filter Problem Formulation}
\label{sec:af_problem}
We define an AF as an algorithm or optimizer that solves 
\begin{equation}
\hat{\btheta}[\tau] = \arg\min_{\btheta[\tau]} \L(\cdots)[\tau] \label{eq:general_af}
\end{equation}
via an additive update rule of the form 
\begin{equation}
\btheta[\tau + 1] = \btheta[\tau] + \bdelta[\tau],
\label{eq:general_update}
\end{equation}
where $\hat\btheta[\tau]$ is a set of estimated time-varying filter parameters. In this work, we focus exclusively on linear FD adaptive filters~(FDAFs), where $\btheta[\tau] = \{ \w[\tau] \}$ without loss of generality.
Common losses include the mean-square error~(MSE),
\begin{equation}
\L_{MSE}[\frame] = E[\|\e_{\mic}[\tau]\|^2] = E[\|\d_{\mic}[\tau] - \y_{\mic}[\frame]\|^2],\label{eq:mse}
\end{equation}
the instantaneous square error~(ISE),
\begin{equation}
\L_{ISE}[\tau] =  \|\e_{\mic}[\tau]\|^2 = \|\d_{\mic}[\tau] - \y_{\mic}[\tau]\|^2,\label{eq:ise}
\end{equation}
and the weighted least squares error~(WSE),
\begin{equation}
\L_{WSE}[\frame] = \sum_{n=0}^{\frame} \gamma^{\frame-n} \|\d_{\mic}[n] - \y_{\mic}[n]\|^2, 
\end{equation}
where $\gamma$ is a forgetting factor, $m$ is a reference mic, and the output $\y_m$ is computed via~\eref{eq:ols_output} or~\eref{eq:ola_freq} for FD signals or losses and~\eref{eq:ols_output_time} or~\eref{eq:ola_output} for time-domain signals or losses. 

\subsection{Conventional Optimizers}
For audio AFs, it is common to leverage OLS or OLA filtering and solve~\eref{eq:general_af} via optimization methods per frequency bin. So, we modify~\eref{eq:general_update} to be
\begin{equation}
\w_{\freq}[\tau + 1] =\w_{\freq}[\tau] + \bdelta_{\freq}[\tau],
\end{equation}
where the update $\bdelta_{\freq}[\tau]$ is per frequency $\freq$. We focus on three common conventional AF optimizers in this form as well as a machine learning optimizer to ground the development of our method to familiar, fundamental algorithms. For gradient methods, we use the partial derivative with respect to $\w_{\freq}[\frame]^{\H}$.

\subsubsection{Least Mean Square}
The least mean square optimizer~(LMS) is a stochastic gradient descent method that uses the ISE loss and gradient. The LMS update is 
\begin{equation}
    \bdelta_{\freq}[\tau] = - \lambda \grad_{\freq}[\frame], \label{eq:lms_update}
\end{equation}
where $\lambda$ is the step-size and $\grad_{\freq}[\frame]$ is the gradient of the ISE. Note, LMS is stateless and only a function of the gradient. 

\subsubsection{Normalized Least Mean Square}
The normalized LMS~(NLMS) algorithm modifies LMS via a running normalizer based on the input power. The NLMS update is 
\begin{align}
    \mathbf{v}_{\freq}[\frame] &= \gamma \mathbf{v}_{\freq}[\frame - 1] + (1 - \gamma) \|
    \u_{\freq}[\frame]\|^2\\
    \bdelta_{\freq}[\tau] & =  - \lambda \frac{\grad_{\freq}[\frame]}{\mathbf{v}_{\freq}[\frame]}, \label{eq:nlms_update}
\end{align}
where the division is element-wise, $\lambda$ is the step-size and $\gamma$ is a forgetting factor. 

\subsubsection{Root Mean Squared Propagation}
The root mean square propagation~(RMSProp) optimizer~\cite{hinton2012neural} modifies NLMS by replacing $\mathbf{v}_{\freq}[\frame]$ with a gradient-based per-element running normalizer, $\boldsymbol{\nu}[\frame]$ using forget factor $\gamma$ as,
\begin{align}
    \boldsymbol{\nu}_{\freq}[\frame] &= \gamma  \boldsymbol{\nu}_{\freq}[\frame - 1] + (1 - \gamma) \|\grad_{\freq}[\frame]\|^2 \\
    \bdelta_{\freq}[\tau] &=  - \lambda \frac{\grad_{\freq}[\frame]}{\sqrt{\boldsymbol{\nu}_{\freq}[\frame]}}, \label{eq:rms_update}
\end{align}
The value, $1 / \sqrt{\boldsymbol{\nu}_{\freq}[\frame]}$ supplements the fixed step-size $\lambda$ and acts as an adaptive learning rate, $\lambda / \sqrt{\boldsymbol{\nu}_{\freq}[\frame]}$.

\subsubsection{Recursive Least Squares}
The aim of the recursive least squares~(RLS) algorithm is to exactly solve the AF loss, most commonly the WSE error. 
This is accomplished by expanding the weighted least squares error into a function of the weighted empirical signal covariance matrix, $\mathbf{\Phi}_{\freq}[\frame] =\sum_{\frame}\gamma^{N-\frame}\u_{\freq}[\frame]\u_{\freq}[\frame]^\H$, where the summation time-indices are application dependent~(e.g. causal vs. non-causal implementations), and the~(weighted) empirical cross-correlation vector $\mathbf{z}_{\freq}[\frame]=\sum_{\frame}\gamma^{N-\frame}\u_{\freq}[\frame]\d_{\freq}[\frame]^\H$ to compute the exact solution to the resulting normal equations, $\mathbf{\Phi}_{\freq}[\frame]\w_{\freq}[\frame] =\mathbf{z}_{\freq}[\frame]$. Running estimates of $\mathbf{\Phi}_{\freq}[\frame]$ and $\mathbf{z}_{\freq}[\frame]$ are also commonly used. However, instead of performing repeated matrix inversion, the matrix inversion lemma is used. Thus, RLS can be implemented using a time-varying precision~(inverse covariance) matrix $\mathbf{P}_{\freq}[\frame]$ and the Kalman-gain $\boldsymbol{\kappa}_{\freq}[\frame]$,
\begin{align}
    \boldsymbol{\kappa}_{\freq}[\frame] &= \frac{\mathbf{P}_{\freq}[\frame - 1] \u_{\freq}[\frame]}{\gamma + \u_{\freq}[\frame]^\H \mathbf{P}_{\freq}[\frame - 1] \u_{\freq}[\frame]}\\
    \mathbf{P}_{\freq}[\frame] &= \frac{\mathbf{P}_{\freq}[\frame - 1]  - \boldsymbol{\kappa}_{\freq}[\frame] \u_{\freq}[\frame]^\H \mathbf{P}_{\freq}[\frame - 1] }{\gamma}\\
    \bdelta_{\freq}[\tau] &= \boldsymbol{\kappa}_{\freq}[\frame](d_{\freq \mic}[\frame] - y_{\freq \mic}[\frame])^*, \label{eq:rls_update}
\end{align}
where $\gamma$ is a forgetting factor, $d_{\freq \mic}[\frame]$ and $y_{\freq \mic}[\frame]$ are the desired and estimated signal at frequency $k$ and reference microphone $\mic$, and the initialization of $\mathbf{P}_{\freq}[\frame]$ is critical and commonly based on the input SNR. 
In the case of multi-block and/or multi-channel filters, there are multiple ways to formulate RLS, some of which differ from time-domain RLS. Common approaches include diagonalized RLS~(D-RLS) and block diagonalized RLS~(BD-RLS) as well as QR decomposition techniques~\cite{alexander1993method, strobach1996low, apolinario2009qrd} and differ in what terms of the covariance~(precision) matrix are modeled. D-RLS makes an independence assumption and optimizes each $\freq,\mic,\mathrm{b}$ filter tap separately, forming $\mathrm{K}$ diagonal $\mathrm{B}\mathrm{M} \times \mathrm{B}\mathrm{M}$ precision matrices. BD-RLS couples across frames and channels by forming $\mathrm{K}$ separate $\mathrm{B}\mathrm{M} \times \mathrm{B}\mathrm{M}$ precision matrices. In the case of single block/channel filters, D-RLS, BD-RLS, and NLMS can be reduced to the same algorithm with different parameterizations. In our case, we use identical BD-RLS implementations across all tasks.

When conventional optimizers are compared to each other, the order of performance commonly follows LMS, RMSProp, NLMS, and RLS, while the order of computational complexity is reversed. These algorithms, however, can be sensitive to tuning, nonstationarities, nonlinearities, and other issues that require engineering effort to mitigate failure cases and stability. For multi-block BD-RLS filters, in particular, poor partition conditioning can also lead to degraded RLS performance~\cite{yang2021convergence} and/or stability issues compared to NLMS and other alternatives. For the purposes of this work, we exhaustively grid-search tune the hyperparemeters on held-out validation sets of signals per task. 

Beyond these basic optimizers, we also compare against several additional methods. For the AEC task, we compare against the double-talk robust diagonal Kalman filter~(D-KF)~\cite{enzner2006frequency}, the open-source double-talk robust Speex algorithm~\cite{valin2007adjusting}, a weighted-RLS~(wRLS) algorithm~\cite{wang2021weighted}, and WebRTC-AEC3~\cite{webrtc_aec3}. The D-KF algorithm is recommended over other variants~\cite{kuech2014state} and is a common AEC baseline~\cite{haubner2021end}. In addition, the Speex and wRLS~\cite{wang2021weighted} algorithms were the linear AFs used~(with a non-linear post-processor) in the first-~\cite{valin2021low} and second-place~\cite{wang2021weighted} winners of the ICASSP 2021 Acoustic Echo Cancellation Challenge~\cite{SpeexIsBest}, respectively. Since our work focuses on linear adaptive filters and can easily be combined with non-linear post-processors, we believe D-KF, wRLS, Speex, and WebRTC-AEC3 are reasonable baselines. For dereverberation, we compare against NARA-WPE~\cite{drude2018nara}, which is a highly effective normalized BD-RLS based optimizer~\cite{yoshioka2012generalization, drude2018nara} and comparable to the original NTT implementation~\cite{manohar2019acoustic}.

\subsection{Related Work}
Initial work on automatically tuning AFs has been explored in incremental delta-bar-delta~\cite{sutton1992adapting}, Autostep~\cite{mahmood2012tuning}, and elsewhere.
Recent machine learning work discusses the idea of using DNNs to learn entirely novel optimizer update rules from scratch~\cite{andrychowicz2016learning, wichrowska2017learned, metz2019understanding, chen2020training}. We take inspiration from this work, but make numerous advances specific to AFs. In particular, past learned optimizers~\cite{andrychowicz2016learning} are element-wise, offline, real-valued, only a function of the gradient, and are trained to optimize general purpose neural networks. In contrast, we design \emph{online} AF optimizers that use multiple input signals, are complex-valued, adapt block FD linear filters, and integrate domain-specific insights to reduce complexity and improve performance~(coupling across channels and time). Moreover, we deploy learned optimizers to solve AF tasks as the end-goal and do not use them to train downstream neural networks. We also note recent work that uses a supervised DNN to control the step-size of a D-KF for AEC~\cite{haubner2021end} and another that uses a supervised DNN to predict both the step-size and a nonlinear reference signal for AEC~\cite{zhang2022deep}. Compared to these, we replace the entire update with a neural network, do not need supervisory signals, and investigate many tasks.

\section{Proposed Method}
\label{sec:method}

\subsection{Problem Formulation}
\label{sec:afformulation}
We formulate AF algorithm design as a meta-learning problem and train neural networks to learn AFs from data, creating meta-learned adaptive filters. This is in contrast to typical AFs that are manually created by human engineers. To do so, we define a learned optimizer, $g_\bphi(\cdot)$, as a neural network with one or more input signals parameterized by weights $\bphi$ that optimizes an AF loss or \textit{optimizee} $\L(h_\btheta(\cdot), \cdots)$ or $\L$ for short, using an additive update rule 
\begin{align}
    \btheta[\frame + 1] &= \btheta[\frame] + g_\bphi(\cdot).
    \label{eq:update}
\end{align}
We then seek an optimal AF optimizer $g_{\hat{\bphi}}$ over dataset $\mathcal{D}$ 
\begin{equation}
\hat{\bphi} = \arg\min_{\bphi} E_\mathcal{D}[\;  \L_M(\; g_\bphi,  \L(h_\btheta(\cdot), \cdots) \; ) \; ], 
\label{eq:meta}
\end{equation}
where $\L_M$ is a functional that defines the meta (or optimizer) loss that is a function of $g_\bphi$ and an AF loss $\L$ with one or more inputs and  filtering function $h_\btheta$ that itself has one or more inputs and parameters $\btheta$. Intuitively, when we solve~\eref{eq:meta}, we learn a network $g_\bphi(\cdot)$ that solves the AF loss $\L$  when applied repeatedly via an additive update. 
 
\subsection{Optimizee Architecture \& Loss}
\label{sec:optimizee}
The optimizee, or the AF loss $\L$ that is optimized via~\eref{eq:update} is a function of the filter or architecture $h_\btheta(\cdot)$. The filter can be any reasonable differentiable filtering operator such as time-domain FIR filters, lattice FIR filters, non-linear filters, FD filters, multi-delayed block FD filters~\cite{soo1990multidelay}, etc.  Similarly, the AF loss can be any reasonable differentiable loss such as the MSE, ISE, WSE, a regularized loss, negative log-likelihood, mutual information, etc.  For our work, we focus on single- and multi-channel multi-frame linear block FD filters $h_\btheta$ applied via OLS or OLA with parameters $\btheta_{\freq}[\frame] = \{\w_{\freq}[\frame] \in \mathbb{C}^{BM} \}$ with $B$ buffered frames and  $M$ channels per frequency $\freq$. We also set the FDAF loss $\L[\frame]$ to be the ISE via~\eref{eq:ise} with gradient computed with respect to $\w_{\freq}[\frame]^{\H}$ as  $\grad_{\freq}[\frame] = \u_{\freq}[\frame](y_{\freq\mic}[\frame] - d_{\freq\mic}[\frame])^*$.
\subsection{Optimizer Architecture \& Loss}
\label{sec:optimizer}

\begin{table}
\begin{center}
\caption{Relationship between AF optimizers.}

\begin{tabular}{ |c|c|c|c|c| } 
\hline
Optimizer & Inputs & State & Params  & $\bdelta_{\freq}[\frame]$ \\
\hline
LMS & $\grad_{\freq}[\frame]$ 
    & $\emptyset$ 
    & $\lambda$ 
    & \eref{eq:lms_update}
    \\
\hline
NLMS & $\grad_{\freq}[\frame], \u_{\freq}[\frame]$ 
    & $\mathbf{v}_{\freq}[\frame]$ 
    & $\lambda, \gamma$ 
    & \eref{eq:nlms_update}
    \\
\hline
RMSProp & $\grad_{\freq}[\frame]$ 
        & $\boldsymbol{\nu}_{\freq}[\frame]$ 
        & $\lambda, \gamma$ 
        &  \eref{eq:rms_update}
        \\
\hline
BD-RLS & $\u_{\freq}[\frame], \d_{\freq}[\frame], \y_{\freq}[\frame]$ 
    & $\mathbf{P}_{\freq}[\frame]$ 
    & $\gamma$ 
    &  \eref{eq:rls_update}
    \\
\hline
Ours & $\bxi_{\freq}[\frame]$ & $\boldsymbol{\psi}_{\freq}[\frame]$ & $\bphi$  & $g_\bphi$ \\
\hline
\end{tabular}
\label{tab:optimizer_relation}
\end{center}
\end{table}

\subsubsection{Architecture}
Our optimizer $g_\bphi$ is inspired by conventional AF optimizers such as LMS, NLMS, and BD-RLS, but updated to have a neural network form. In particular, we focus on making a generalized, stochastic variant of an RLS-like optimizer that is applied independently per frequency $\freq$ to our optimizee parameters, but coupled across channels and time frames to allow our approach to model interactions between channels and frames and vectorize across frequency. To do so, we use a recurrent neural network~(RNN) where the weights $\bphi$ are shared across all frequency bins, but we maintain separate state $\boldsymbol{\psi}_{\freq}[\frame]$ per frequency. The inputs to our optimizer at frequency $\freq$ are $\bxi_{\freq}[\frame] = [\grad_{\freq}[\frame], \u_{\freq}[\frame], \d_{\freq}[\frame], \y_{\freq}[\frame], \e_{\freq}[\frame]]$, where $\grad_{\freq}[\frame]$ is the gradient of the optimizee with respect to $\btheta_\freq$, and $\e_{\freq}[\frame] = \d_{\freq}[\frame] - \y_{\freq}[\frame]$. Our optimizer outputs are the update $\bdelta_{\freq}[\frame]$ and the  internal state $\boldsymbol{\psi}_{\freq}[\frame+1]$, resulting in 
\begin{eqnarray}
(\bdelta_{\freq}[\frame], \boldsymbol{\psi}_{\freq}[\frame+1]) &=& g_\bphi(\bxi_{\freq}[\frame], \boldsymbol{\psi}_{\freq}[\frame]) \\
\btheta_{\freq}[\frame + 1] &=& \btheta_{\freq}[\frame] +  \bdelta_{\freq}[\frame].
\end{eqnarray}
Our design is in contrast to LMS-, NLMS-, RMSProp-like optimizers that have no state~(e.g. LMS) or minimal state dynamics~(e.g. NLMS, RMSProp). In addition, these optimizers as well as other learned optimizers~\cite{andrychowicz2016learning} typically apply updates independently per element. For a comparison of optimizer inputs, state, parameters, and gradients, please see~\tref{tab:optimizer_relation}. 

To define our RNN in more detail, we use a small network composed of a linear layer, nonlinearity, and two Gated Recurrent Unit~(GRU) layers with hidden size $H=32$, followed by two additional linear layers with nonlinearities, where all layers are complex-valued. We always re-scale the inputs element-wise via
\begin{equation}
    \ln(1 + |\bxi|)e^{j\angle\bxi}, \label{eq:whiten}
\end{equation}
\noindent to reduce the dynamic range and facilitate training, but keep the phases unchanged. This pre-processing was found useful in several previous works~\cite{casebeer2021auto, andrychowicz2016learning}, although previous work used explicit clipping, which we found unnecessary.

\subsubsection{Loss}
We examine two meta losses $\L_M(\cdot)$ to learn our optimizer parameters $\bphi$. 
First, we define the FD \emph{frame independent} loss 
\begin{eqnarray}
\L_M &=& \ln \frac{1}{L}\sum_{\frame}^{\tau+L} E[ ||\d_\mic[\tau] - \y_\mic[\tau]||^2],
\label{eq:oloss1}
\end{eqnarray}
where $\d_\mic[\frame]$ and $\y_\mic[\frame]$ are the desired and estimated FD signal vectors of the reference channel $\mic$ (e.g. $\mic=0$). Intuitively, to compute this loss for a given optimizer $g_\bphi$, we unroll~\eref{eq:update} for a time horizon of $L$ time frames, compute the FD mean-squared error over $L$ frames, and then take the logarithm to reduce the dynamic range, which we found to empirically improve learning. This loss ignores the temporal order of AF updates and optimizes for filter coefficients that are unaware of any downstream STFT processing, but the idea of  accumulating independent time-step losses is common~\cite{andrychowicz2016learning}.

Second, we define the time-domain \emph{frame-accumulated} loss
\begin{eqnarray}
\L_M \hspace{-.2cm} &=& 
     \hspace{-.2cm} \ln E[||\bar{\d}_\mic[\frame] - \bar{\y}_\mic[\frame]||^2] \label{eq:oloss2} \\
 \bar{\d}_\mic[\tau] \hspace{-.2cm} &=&  
    \hspace{-.2cm}\tiny{\mathrm{cat}(\underline\d_\mic[\tau], \underline\d_\mic[\tau+1],\tiny{\cdots},\underline\d_\mic[\tau+L-1])} \\
\bar{\y}_\mic[\tau] \hspace{-.2cm} &=& 
    \hspace{-.2cm} \tiny{\mathrm{cat}(\underline\y_\mic[\tau], \underline\y_\mic[\tau+1], \tiny{\cdots},\underline\y_\mic[\tau+L-1])},
\end{eqnarray}
where $\underline\d_\mic[\frame]$ and $\underline\y_\mic[\frame]$ are the time-domain desired and estimated responses of reference channel $\mic$ and $\bar{\d}_\mic[\tau] \in R^{RL}$ and $\bar{\y}_\mic[\tau] \in R^{RL}$.  Intuitively, to compute this loss for a given optimizer $g_\bphi$, we run~\eref{eq:update} for a time horizon of $L$ frames, concatenate the sequence of time-domain outputs and target signals to form longer signals, then compute the time-domain MSE loss, and take the logarithm. While both losses use the same time-horizon, the frame accumulated loss allows us to model boundaries between adjacent updates and implicitly learn updates that are STFT consistent~\cite{le2008explicit}. To the best of our knowledge, our frame accumulated loss is novel for AFs. 

\subsubsection{Computational Complexity}
We compare the computational complexity of our proposed approach to conventional optimizers in~\tref{tab:optimizer_complexity}. We note that the complexity of our approach is dependent on the hidden state size $\mathrm{H}$ of our RNN, but is linear in channels $\mathrm{M}$ and blocks $\mathrm{B}$, whereas BD-RLS is quadratic in $\mathrm{M}$ and $\mathrm{B}$, but does not depend on $\mathrm{H}$. Note that prior work on learned optimizers, including our own~\cite{casebeer2021auto}, performs optimization completely element-wise, resulting in a larger complexity of $\mathcal{O}(\mathrm{KMBH^2})$. 

\begin{table}
\begin{center}
\caption{Optimizer complexity comparison.}
\begin{tabular}{ |c|c|c| } 
\hline
Optimizer & Big-$\mathcal{O}$ & $\approx \mathbb{C}$MACS \\
\hline
LMS & $\mathcal{O}(\mathrm{KMB})$
    & $\mathrm{KMB}$
    \\
\hline
NLMS & $\mathcal{O}(\mathrm{KMB})$
    & $5\mathrm{KMB}$
    \\
\hline
RMSProp & $\mathcal{O}(\mathrm{KMB})$
        & $6\mathrm{KMB}$
        \\
\hline
BD-RLS & $\mathcal{O}(\mathrm{K(MB)^2})$
    & $\mathrm{K}(4 (\mathrm{MB})^2 + 5 \mathrm{MB})$
    \\
\hline
Ours & $\mathcal{O}(\mathrm{K(\mathrm{H}^2 + \mathrm{MBH})})$
    & $\mathrm{K}(12\mathrm{H}^2 + (21 + 10\mathrm{MB})\mathrm{H})$
    \\
\hline
\end{tabular}
\label{tab:optimizer_complexity}
\end{center}
\end{table}

\subsection{Learning the Optimizer}
To learn an optimizer $g_\bphi$ from data, we solve~\eref{eq:meta} using standard deep learning tools~(i.e. JAX~\cite{bradbury2020jax, haiku2020github}), including the use of automatic differentiation for training and inference. In addition, we use truncated backpropagation through time~(TBPTT)~\cite{werbos1990backpropagation} with a stochastic gradient descent optimizer, Adam~\cite{kingma2014adam}, that we call our meta optimizer. We show a simplified form of our training algorithm in Alg.~\ref{alg:inner_outer} using our frame accumulated loss and a batch size of one, where \textproc{Stft} is an OLA or OLS processor, \textproc{Grad} returns the gradient of the first argument with respect to the second, \textproc{Sample} randomly samples signals from a dataset $\mathcal{D}$, and \textproc{NextL} grabs the next $L$ time buffers from a longer signal. In practice, we use batching.

\begin{algorithm}[t]
  \caption{Training algorithm.}
  \begin{algorithmic}
    \Function{Forward}{$g_\bphi, \boldsymbol{\psi}, h_\btheta, \underline\U, \underline\d_\mic$}
      \For{$\frame \gets 0 \textrm{ to } L $}
      \Comment{Unroll}
        
        \Let{$\U[\frame], \d[\frame]$}{\textproc{Stft}$(\underline\U[\frame], \underline\d_{\mic}[\frame])$}
        \Comment{Forward STFT}
      
        \Let{$\y_{\mic}[\frame]$}{$h_{\btheta}(\U[\frame])$}
        \Comment{Save filter output}
        
        \Let{$\underline\y_\mic[\frame]$}{\textproc{Stft}$^{-1}(\y_\mic[\frame])$}
        \Comment{Inverse STFT}
        
        \Let{$\L$}{$||\d_{\mic}[\tau] - \y_{\mic}[\tau]||^2$}
        \Comment{AF frame loss}
        
        \Let{$\grad[\frame]$}{\textproc{Grad}$(\L, \btheta$)}
        \Comment{Filter gradient}
        
        \For{$\freq \gets 0 \textrm{ to } K$}
          \Comment{Apply update per freq}
          \Let{$\bxi_{\freq}[\frame]$}{$[\grad_{\freq}[\frame], \u_{\freq}[\frame], \d_{\freq}[\frame], \y_{\freq}[\frame],
          \e_{\freq}[\frame]]$}
          
          \Let{$(\bdelta_{\freq}[\frame], \boldsymbol{\psi}_{\freq}[\frame+1])$}{$g_\bphi(\bxi_{\freq}[\frame], \boldsymbol{\psi}_{\freq}[\frame])$}
          
          \Let{$\btheta_{\freq}[\frame + 1]$}{$\btheta_{\freq}[\frame] +  \bdelta_{\freq}[\frame]$}
        \EndFor
      \EndFor

         \Let{$\bar{\y}$}{\textproc{Cat}$(\underline\y[\frame] \; \forall \frame)$}
         \Comment{Concatenate accumulated frames}
 
      \State \Return{$\bar{\y}, \boldsymbol{\psi}, h_\btheta$}
    \EndFunction
 
    \Function{Train}{$\mathcal{D}$}
      \Let{$\bphi$}{\cite{wolter2018complex} init}
      \While{$\bphi$ \textbf{not} \textproc{Converged}}
      \Comment{Train loop}
          \Let{$\underline\U, \underline\d_\mic$}{\textproc{Sample}($\mathcal{D}$)}
          \Comment{Sample signals}
          \Let{$\btheta, \boldsymbol{\psi}$}{$\mathbf{0}, \mathbf{0}$}
          \Comment{Init filter and optimizer state}
          \For{$n \gets 0 \textrm{ to end }$}
          \Comment{Loop across long signal}
          \Let{$\bar\U, \bar\d_\mic$}{\textproc{NextL}$(\underline\U, \underline\d_\mic)$}
          \Comment{Get next $L$ frames}
          \Let{$\bar{\y}, \boldsymbol{\psi}, h_\btheta$}{\textproc{Forward}($g_\bphi, \boldsymbol{\psi}, h_\btheta, \bar\U, \bar\d_\mic$)}
        
          \Let{$\L_M$}{via~\eref{eq:oloss2}}
          \Comment{Meta loss}
         
          \Let{$\grad$}{\textproc{Grad}($\L_M$, $\bphi$)}
          \Comment{Optimizer gradient}
          \Let{$\bphi[n + 1]$}{\textproc{MetaOpt}($\bphi[n], \grad$)}
          \Comment{Update opt}
          \EndFor
      \EndWhile
      \Return{$\hat{\bphi}$}
      \Comment{Return best $\bphi$}
    \EndFunction
  \end{algorithmic}
  \label{alg:inner_outer}
\end{algorithm}


In more detail, we train $g_\bphi$ until the application specific mean SNR metric on the validation fold of a dataset $\mathcal{D}$ has not improved for four epochs. For each of our five applications, we use ~\eref{eq:id_snr},~\eref{eq:aec_snr},~\eref{eq:eq_snr1},~\eref{eq:wpe_snr}, and~\eref{eq:gsc_snr}, respectively.
We halve the step size after an epoch with no improvement and define an epoch as $10$ passes through the dataset with a batch size of $32$. We have a hard stop for training at $100$ epochs. 
From initial experimentation, we note the choice of the meta-optimizer and meta-optimizer parameters have large impact on performance.
We initialize $\bphi$ via~\cite{wolter2018complex} and, when using Adam\footnote{We use a custom implementation of Adam in our source code due to a complex-value error issue in the JAX implementation.}, we found it was important to use a small learning of $10^{-4}$ and a large momentum term of $.99$. 
To help stabilize training, we use gradient clipping with a clipping value of $10$.
We use identical $g_\bphi$ settings for each task with input and output sizes set to $M \cdot B \cdot 5$ and $M \cdot B$, respectively, all intermediate layer sizes set to $32$, and  separate ReLU nonlinearities for real and imaginary.  

Compared to our past work on single-channel single-talk AEC~\cite{casebeer2021auto}, we 1) change our loss from~\eref{eq:oloss1} without the log to~\eref{eq:oloss2} 2) change the input of our learned optimizer from $\grad_{\freq}[\frame]$ to $\bxi_{\freq}[\frame]$ with updated pre-processing~\eref{eq:whiten} and 3) develop multi-channel Meta-AFs or development multi-block Meta-AFs to perform time- and block-coupled updates instead of separate, independent updates per frequency, block, and channel. To validate our approach, we apply our algorithm to five audio tasks including system identification, acoustic echo cancellation, equalization, single/multi-channel dereverberation, and beamforming in~\sref{sec:sys},~\sref{sec:aec},~\sref{sec:eq},~\sref{sec:dereverb}, and~\sref{sec:beamforming}, respectively and compare against conventional methods. For our first task, we also ablate key design decisions and then lock a single configuration for all remaining tasks and experiments.

\section{System Identification Ablations}
\label{sec:sys}

\begin{figure}[!t]
    \centering
    \includegraphics[width=.97\linewidth]{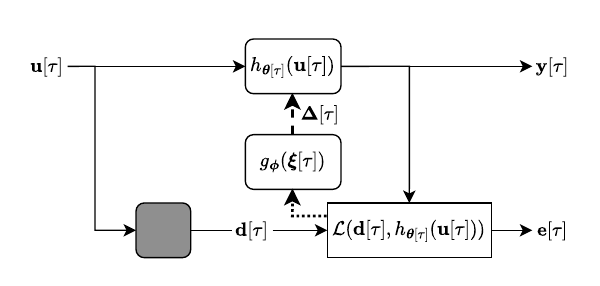}
    \vspace{-5mm}
    \caption{System identification block diagram. System inputs are fed to both the adaptive filter and the true system~(shaded box). The adaptive filter is updated to mimic the true system.}
    \label{fig:sys}
\end{figure}

\subsection{Problem Formulation}
For our first task, we train a Meta-AF to perform online system identification~(ID). We seek to estimate the transfer function between an audio source and a microphone over time, as shown in~\fref{fig:sys}. This is commonly done in room acoustics and head-related transfer function measurements for virtual and augmented reality. To do so, we model the unknown system~(e.g. acoustic room) with a linear frequency-domain filter $h_\btheta$~(optimizee architecture), inject input signal $\u$ into the system, measure the response $\d$, and adapt the filter weights $\btheta=\{\w_{\freq}\}$ using our learned Meta-ID AF, $g_\bphi$. The AF loss is the ISE between the desired response, $\d_{\freq}[\frame]$, and the AF predicted response, $\y_{\freq}[\frame]=\w_{\freq}[\frame]^\H \u_{\freq}[\frame]$.

\subsection{Experimental Design}
We 1) ablate key design decisions of our AF optimizer architecture and loss in~\sref{sec:design_decisions} 
and 2) study the robustness of our approach to modeling errors by changing the AF and true system order at test time in~\sref{sec:system_order}. 
For evaluation metrics, we use the segmental SNR~($\operatorname{SNR}_d$) between the desired and estimated signals in dB. We compute this per-frame as,
\begin{equation}
    \operatorname{SNR}_d(\underline{\d}[\frame], \underline{\y}[\frame]) = 10 \cdot \log_{10} \left(\frac{\|\underline{\d}[\frame]\|^2}{\|\underline{\d}[\frame] - \underline{\y}[\frame]\|^2} \right),
    \label{eq:id_snr}
\end{equation}
where higher is better. 
We train $g_\phi$ via~\aref{alg:inner_outer} using a single GPU to adapt an OLS filter with a rectangular window size $N=2048$ and hop size $R=1024$ on $16\,$kHz audio. 
For training data, we created a dataset by convolving the far-end speech from the single-talk portion of the ICASSP 2021 AEC Challenge data~\cite{cutler2022AEC} with room impulse responses~(RIRs) from~\cite{ko2017study} truncated to $1024$ taps. 

\subsection{Optimizer Design Results \& Discussion}
\label{sec:design_decisions}
\begin{figure}[!t]
    \centering
    \includegraphics[width=.97\linewidth]{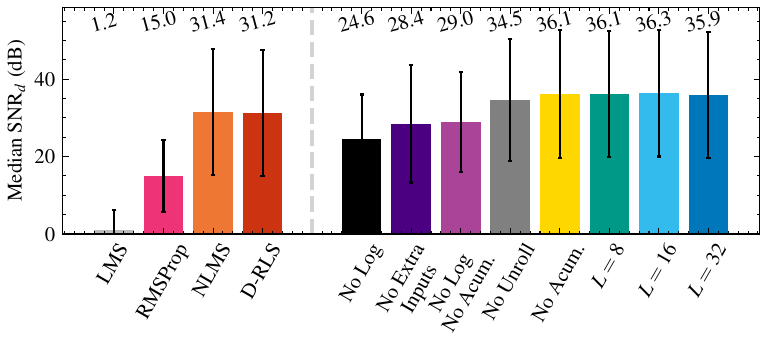}
    \caption{Optimizer design decision ablation. Using an accumulated log-loss leads to our best model, particularly for more complex tasks we address later on. RLS-like inputs are also useful. The exact value of $L$ is not critical, but larger is better.}
    \label{fig:mainablation}
\end{figure}

In~\fref{fig:mainablation}, we change one aspect of our final design at a time and show how each change negatively effects performance. 
Our final design, $L=16$, is colored light blue, alternative configurations are colored differently, and our previous work\cite{casebeer2021auto} is approximately equivalent to a no log, no accumulation, no extra inputs setting.
After this ablation, we fix these values for all remaining experiments and do not perform any further tuning except changing the dataset used for training and using different checkpoints caused by early stopping on validation performance. In contrast, we re-tune all conventional optimizer baselines for all subsequent tasks on held-out validation sets.

\subsubsection{Loss Function}
First, we compare our selected frame accumulated loss model~(light blue) to the frame accumulated loss without log scaling~(black) as well as the frame independent loss~(yellow) and without log scaling~(light-purple). As shown, the log-loss has the single largest effect on $\operatorname{SNR}_d$ and yields an astounding $11.7/7.3\,$dB improvement compared to the no-log losses. When we compare the independent vs. accumulated loss, the accumulation provides a $.2\,$dB improvement. However, when we listen to the estimated response, especially in more complex tasks, we found that the accumulated loss introduces fewer artifacts and perceptually sounds better. Thus, we fix the optimizer loss to be~\eref{eq:oloss2}. 

\subsubsection{Input Features}
Having selected the optimizer loss, we compare the model inputs. We compare setting the optimizer input to be only the gradient $\bxi_{\freq}[\frame] = [\grad{\freq}[\frame]]$ for an LMS-like learned optimizer~(dark purple) and setting the optimizer to be the full signal set $\bxi_{\freq}[\frame] = [\grad{\freq}[\frame], \u_{\freq}[\frame], \d_{\freq}[\frame], \y_{\freq}[\frame], \e_{\freq}[\frame]]$ for our selected RLS-like learned optimizer~(light blue). As shown, the inputs have the second largest effect on $\operatorname{SNR}_d$ and using the full signal set yields a notable $7.9\,$dB improvement. Thus, we set the input to be the full signal set.

\subsubsection{AF Unroll}
With the optimizer loss and inputs fixed, we evaluate four different values of AF unroll length, $L=2, 8, 16, 32$, where $L$ is the number of time-steps over which the optimizer loss is computed in~\eref{eq:oloss2}. Intuitively, a larger unroll introduces less truncation bias but may be more unstable during training due to exploding or vanishing gradients. The case where $L=2$ corresponds to no unroll, since for $L=1$ the meta loss is not a function of the optimizer parameters and yields a zero gradient. As shown, for no unroll $L=2$~(grey), we get a reduction of the $\operatorname{SNR}_d$ by $1.8\,$dB compared to our best model. When selecting the unroll between $8, 16, 32$, however, there is a small~($<1\,$dB) overall effect. That said, we find that longer unroll values work better in challenging scenarios but take longer to train. As a result, we select an unroll length of $16$, as it represents a good trade-off between performance and training time. Note, the unroll length only effects training and is not used at test time.

\subsection{System Order Modeling Error Results \& Discussion}
\label{sec:system_order}

\begin{figure}[!t]
    \centering
    \includegraphics[width=.97\linewidth]{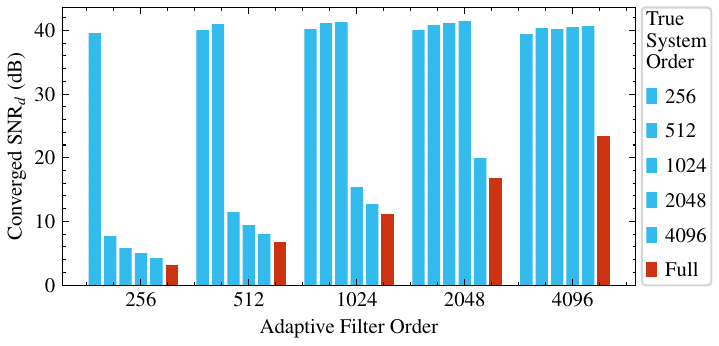}
    \caption{Evaluating the effect of different true system orders and adaptive filter orders. Our learned AFs can operate well across a variety of linear system orders, even when training is restricted to systems of a fixed length~($1024$ taps).}
    \label{fig:syslength}
\end{figure}
Given our fixed set of optimizer and meta-optimizer parameters, we evaluate the robustness of our Meta-ID AF to modeling errors by studying what happens when we use an optimizee filter that is too short or too long compared to the true system. We do so by testing a learned optimizer with 1) optimizee filter lengths between $256$ and $4096$ taps and 2) held-out signals with true filter lengths between $256$ and $4096$, as well as full length systems~(up to $32,000$ taps).

Results are shown in~\fref{fig:syslength}. We measure performance by computing the segmental $\operatorname{SNR}_d$ score via~\eref{eq:id_snr} at convergence.
As expected, when the adaptive filter order is equal to or greater than the true system order, we achieve SNRs of $\approx40\,$dB. It is interesting to note that our learned AF was only ever trained on optimizee filters with an order of $1024$ and $1024$ tap true systems. This experiment suggests our learned optimizers can generalize to new optimizee filter orders.

\section{Acoustic Echo Cancellation Ablation}
\label{sec:aec}

\subsection{Problem Formulation}
\begin{figure}[!t]
    \centering
    \includegraphics[width=.97\linewidth]{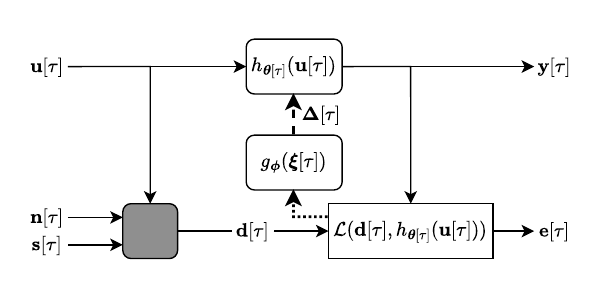}
    \vspace{-5mm}
    \caption{AEC block diagram. System inputs are fed to the adaptive filter and true system~(shaded box). The adaptive filter is updated to mimic the true system. The desired response can be noisy due to near-end noise and speech~($\mathbf{n}[\frame], \mathbf{s}[\frame]$).}
    \label{fig:aec}
\end{figure}

In our second task, we train a Meta-AF for acoustic echo cancellation~(AEC). The goal of AEC is to remove the far-end echo from a near-end signal for voice communication by mimicking a time-varying transfer function as show in~\fref{fig:aec}. The far-end refers to the signal transmitted to a local listener and the near-end is captured by a local mic. We model the unknown transfer function between the speaker and mic with a linear multi-delay frequency-domain filter $h_{\btheta}$, measure the noisy response $\d$ which includes the input signal $\u$, noise $\mathbf{n}$, and near-end speech $\s$, and adapt the filter weights $\btheta$ using our learned Meta-AEC AF, $g_{\bphi}$. The time-domain signal model is $\underline{\d}[\time]=\underline{\u}[\time]\ast\underline{\w}+\underline{\mathbf{n}}[\time]+\underline{\mathbf{s}}[\time]$. The AF loss is the ISE between the near-end and the predicted response. The FDAF near-end speech estimate is $\hat{\s}_{\freq}[\frame]=\d_{\freq}[\frame] - \w_{\freq}[\frame]^\H \u_{\freq}[\frame]$.

\subsection{Experimental Design}
We compare our approach to LMS, RMSProp, NLMS, BD-RLS, a diagonal Kalman filter~(D-KF) model~\cite{enzner2006frequency}, and Speex~\cite{valin2007adjusting} for a variety of acoustic echo cancellation scenarios. Our scenarios, in increasing difficulty, include single-talk, double-talk, double-talk with a path change, and noisy double-talk with a path change and non-linearity. Single-talk refers to the case where only the far-end input signal $\u$ is active. Double-talk refers to the case where both the far-end signal $\u$  and near-end talker signal $\s$ are active at the same time. 

A path change refers to the case where the true system transfer function is abruptly changed (e.g. during a phone call). Nonlinearities refer to the case where the true system is not strictly linear (e.g. harmonic loudspeaker distortion). We train a single $g_\bphi$ for AEC and then test it for each scene type against all hyperparameter-tuned baselines.

We measure echo cancellation performance using segmental echo-return loss enhancement~($\operatorname{ERLE}$)~\cite{enzner2014acoustic} and short-time objective intelligibility~($\operatorname{STOI}$)~\cite{taal2011algorithm}. Segmental $\operatorname{ERLE}$ is
\begin{equation}
    \operatorname{ERLE}(\underline{\d}_{\u}[\frame], \underline{\y}[\frame]) = 10 \cdot \log_{10} \left(\frac{\|\underline{\d}_{\u}[\frame]\|^2}{\|\underline{\d}_{\u}[\frame] - \underline{\y}[\frame]\|^2} \right),\label{eq:aec_snr}
\end{equation}
where $\underline{\d}_{\u}$ is the noiseless system response and higher values are better. When averaging, we discard silent frames using an energy-threshold VAD. In scenes with near-end speech, we use $\operatorname{STOI}\in[0, 1]$ to measure the preservation of near-end speech quality. Higher $\operatorname{STOI}$ values are better. We train $g_\bphi$ via~\aref{alg:inner_outer} using one GPU, which took $\approx 72$ hours. We use a four-block multi-delay OLS filter (MDF) with window sizes of $N=1024$ samples and a hop of $R=512$ samples on $16\,$kHz audio. In double-talk scenarios, we use the noisy near-end, $\d$ as the target and do not use oracle  cancellation results~(such as $\underline{\d}_{\u}$). 

With respect to datasets for single-talk, double-talk, and double-talk with path-change experiments, we re-mix the synthetic fold of the  ICASSP 2021 AEC Challenge dataset (ICASSP-2021-AEC)~\cite{cutler2022AEC} with impulse responses from~\cite{ko2017study}. We partition~\cite{ko2017study} into non-overlapping train, test, and validation folds and set the signal-to-echo-ratio randomly between $[-10,10]$ with uniform distribution. To simulate a scene change, we splice two files such that the change occurs randomly between seconds four and six. For the noisy double-talk with nonlinearity experiments, we use the synthetic fold of~\cite{cutler2022AEC}. We apply a random circular shift and random scale to all files, each ten seconds long. For each task, there are $9000$ training, $500$ validation, and $500$ test files. Finally, we also use an unmodified version of the ICASSP-2021-AEC training, validation, and test set~(does not include scene changes) to compare to other previously published works directly.

\subsection{Results \& Discussion}
Overall, we find that our approach significantly outperforms all previous methods in all scenarios, but has a larger advantage in harder scenes---more details discussed below. 

\subsubsection{Single-Talk}
\begin{figure}[!t]
    \centering
    \includegraphics[width=.97\linewidth]{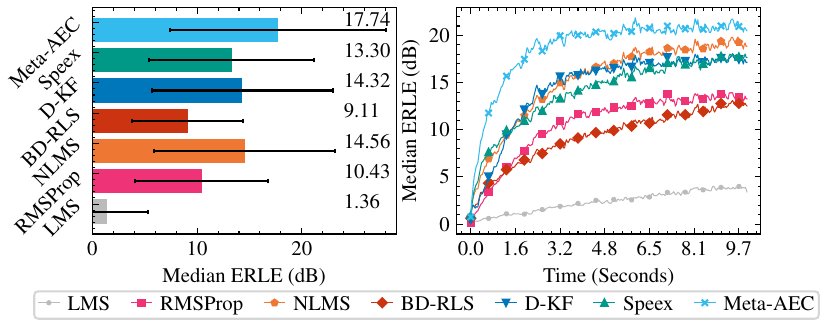}
    \caption{AEC single-talk performance. Meta-AEC converges rapidly and has better steady-state performance. We use this same legend for all AEC plots.}
    \label{fig:aecst}
\end{figure}
Our approach~(light blue, x) exhibits strong single-talk performance and surpasses all baselines by $>\approx3\,$dB in both median and converged $\operatorname{ERLE}$, as shown in~\fref{fig:aecst}. Additionally, Meta-AEC converges fastest, reaching steady-state $\approx4$ seconds before other baselines.

\subsubsection{Double-Talk}
\begin{figure}[!t]
    \centering
    \includegraphics[width=.97\linewidth]{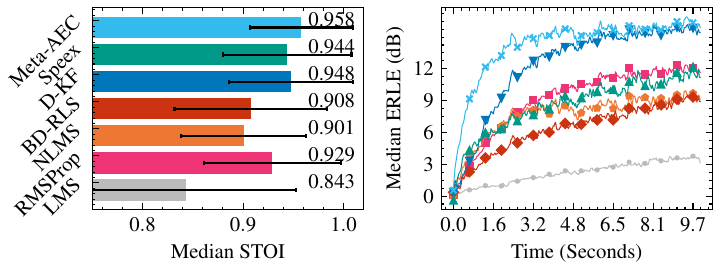}
    \caption{AEC double-talk performance. Meta-AEC converges fastest and has similar peak performance to the D-KF, while preserving near-end speech quality.}
    \label{fig:aecdt}
\end{figure}

Our method~(light blue, x) converges fastest in double-talk, and matches the D-KF in converged-performance, as shown in~\fref{fig:aecdt}. Meta-AEC converges $\approx5\,$ seconds faster while scoring better in $\operatorname{STOI}$. This result is striking as it is typically necessary to either explicitly freeze adaptation via double-talk detectors or implicitly freeze adaption via carefully derived updates as found in both the D-KF model~(dark blue, down triangle) and Speex~(green, up triangle). We hypothesize our method automatically learns how to adapt in double-talk in a completely autonomous fashion.


\subsubsection{Double-Talk with Path Change}
\begin{figure}[!t]
    \centering
    \includegraphics[width=.97\linewidth]{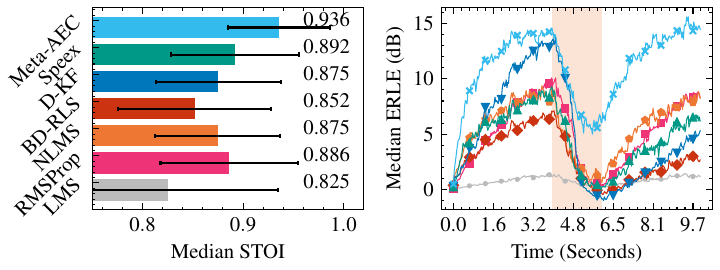}
    \caption{AEC double-talk with a path change~(shaded region) performance. Our approach re-converges rapidly with high speech quality.}
    \label{fig:aecsc}
\end{figure}
Our method~(light blue, x) is able to more robustly handle double-talk with path changes compared to other methods as shown in~\fref{fig:aecsc}. Similar to straight double-talk, our approach effectively learns how to deal with adverse conditions~(i.e. a path change) without explicit supervision, converging and reconverging in $\approx2.5$ seconds, with $.044$ better median $\operatorname{STOI}$. All other algorithms similarly struggle, even Speex~(green, up triangle), which  has explicit self-resetting and dual-filter logic.

\subsubsection{Noisy Double-Talk with Nonlinearities \& Path Change}
\begin{figure}[!t]
    \centering
    \includegraphics[width=.97\linewidth]{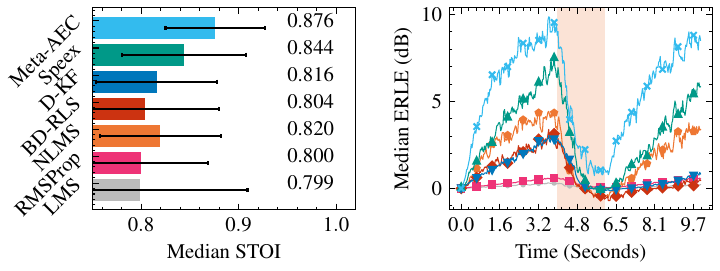}
    \caption{AEC noisy double-talk with nonlinearities and a path change~(shaded region) performance. Meta-AEC learns to compensate for the nonlinearity.}
    \label{fig:aecnl}
\end{figure}
When we evaluate scenes with noise, nonlinearies to simulate loudspeaker distortion, and path changes, we find that our Meta-AEC approach~(light blue, x) continues to perform well, as shown in~\fref{fig:aecnl}. That is, our peak-performance is $\approx2\,$dB above the nearest baseline. In $\operatorname{STOI}$, Meta-AEC scores $.027$ above Speex~(green, up triangle). We hypothesize that our approach effectively learns to  compensate for the signal model inaccuracy, even if we only use a linear filter. 

\subsubsection{ICASSP 2021 AEC Challenge Results}
In addition to testing with our own variant of the ICASSP-AEC-2021 dataset that includes scene changes, we test our work with an unmodified version of the test set in~\tref{tab:ms_aec_benchmark}. Furthermore, we evaluate performance when we train (or tune) on our custom training dataset versus when we train on the original training dataset (denoted with $^\star$). See also a similar Table in past work~\cite{wang2021weighted}. To the best of our knowledge, this dataset is the most recent and widely used dataset for benchmarking AEC algorithms. Here, results from WebRTC-AEC3 and wRLS, $\beta = 0.2$ come from past work~\cite{wang2021weighted}. All other methods have the same MDF filtering architecture as described above.
\begin{table}[t]
    \centering
    \begin{tabular*}{.99\linewidth}{@{\extracolsep{\fill}}c c c c c@{}}
         \toprule
         \textbf{Method} & \textbf{STOI} & \textbf{STOI}$^\star$ & \textbf{ERLE}~(dB) & \textbf{ERLE}$^\star$~(dB)\\
         \bottomrule
         LMS     & 0.794 & 0.794 & 0.937 & 0.560\\
         RMSProp &  0.802 & 0.856 & 1.02 & 4.63 \\
         NLMS    & 0.854 & 0.861 & 4.81 & 5.96 \\
         BD-RLS    & 0.829 & 0.835 & 3.66 & 4.07 \\
         D-KF~\cite{enzner2006frequency} & 0.817 & 0.875 & 1.98 & 6.55 \\
         wRLS, $\beta = 0.2$~\cite{wang2021weighted} & N.A & 0.85 & N.A & N.A \\
         WebRTC-AEC3~\cite{webrtc_aec3} & 0.82 & N.A & N.A & N.A \\
         Speex~\cite{valin2007adjusting}   & 0.869 & N.A & 3.92 & N.A \\
         Meta-AEC & \textbf{0.881} &  \textbf{0.899} &  \textbf{7.73} &  \textbf{9.13}\\
         \bottomrule
    \end{tabular*}
    \caption{ICASSP-2021-AEC test set linear filter results. Our proposed method out performs several past comparable linear-filtering approaches. A $\star$ denotes results when models were trained/tuned on the ICASSP-2021-AEC data.}
    \label{tab:ms_aec_benchmark}
\end{table}
Our approach outperforms all methods we compare against for both training datasets, including Speex and wRLS, which were the linear filters used in the first- and second-place winners of the ICASSP 2021 AEC Challenge~\cite{SpeexIsBest}. Interestingly, there is a significant effect on training or tuning with data that includes scene changes (ours) vs. the original data (e.g. RMSProp and D-KF~\cite{haubner2021end}). That is, because the ICASSP-2021-AEC train and test set does not include scene changes, most algorithms give better performance when trained/tuned on the matching, unmodified train set, even though such results are less realistic.

\subsubsection{Computational Complexity}
Our learned AF has a single CPU core real-time factor~(RTF)~(computation/time) of $\approx0.36$, and $32\,$ms latency (OLS hop size). Our optimizer network alone has $\approx14K$ complex-valued parameters and single CPU core RTF of $\approx0.31$. While this performance is already real-time capable, we suspect it could easily be improved with better optimized code.
\section{Equalization Ablation}
\label{sec:eq}

\subsection{Problem Formulation}
\begin{figure}[!t]
    \centering
    \includegraphics[width=.97\linewidth]{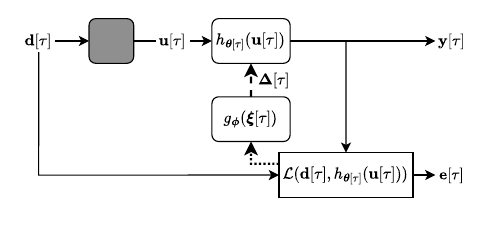}
    \vspace{-5mm}
    \caption{Inverse modeling block diagram. System outputs are fed to the adaptive filter. The adaptive filter is continually updated to invert the unknown system~(shaded box).}
    \label{fig:eq}
\end{figure}

For our third task, we train a Meta-AF for the inverse modeling task of equalization~(EQ). Here, our goal is to estimate the inverse of an unknown transfer function, while only observing input and outputs of the forward system, as shown in~\fref{fig:eq}. This is a common component of loudspeaker tuning. We model the unknown inverse transfer function with a linear frequency-domain filter $h_{\btheta}$, measure the response $\d$ to an input signal $\u$, and adapt the filter weights $\btheta$ using our learned Meta-EQ AF $g_{\bphi}$. The AF loss is the ISE between the true and predicted responses. More precisely, the frequency-domain AF output is $\y_{\freq}[\frame]=\w_{\freq}[\frame]^\H \u_{\freq}[\frame]$.

\subsection{Experimental Design}
We compare our Meta-EQ approach to LMS, RMSProp, NLMS, and D-RLS on the task of frequency equalization. Additionally, we ablate the equalization filtering mechanics for two cases: constrained and unconstrained filters (optimizee architecture modifications). In the constrained case, we set $h_{\btheta}$ to use standard OLS. However, in the unconstrained case, we set $h_{\btheta}$ to use aliased OLS where $\mathbf{Z}_w=\I_K$. This comparison lets us test if Meta-EQ is automatically learning constraint-aware update rules. We train a new $g_\bphi$ for each case~(no separate tuning) and tune all baselines for each case.

We measure performance with signal $\operatorname{SNR}_d$, and system $\operatorname{SNR}_w$. We define these as 
\begin{align}
    \operatorname{SNR}_d(\underline{\d}, \underline{\y}) &= 10 \cdot \log_{10} \left(\frac{\|\underline{\d}\|^2}{\|\underline{\d} - \underline{\y}\|^2} \right) \label{eq:eq_snr1}\\
    \operatorname{SNR}_w(\hat{\w}^{-1}, \w^{-1}) &= 10 \cdot \log_{10} \left(\frac{\||\w^{-1}|\|^2}{\||\w^{-1}| - |\hat{\w}^{-1}|\|^2} \right) \label{eq:eq_snr2}
\end{align}
respectively, where higher is better. We compute $\operatorname{SNR}_w$ using the inverse system magnitude, which ignores the phase. We train $g_\bphi$ via~\aref{alg:inner_outer} on one GPU, which took at most $36\,$h. We use an OLS filter with a window size of $N=1024$ samples and a hop of $R=512$ samples on $16\,$kHz audio.

To construct the equalization dataset, we use speech from the DAPS dataset~\cite{mysore2014can}, take the \emph{cleanraw} recordings as inputs, and apply random equalizer filters from the sox library to generate the outputs, where we randomly pick between  $[5, 15]$ filters with settings $c\in[1,8]\,$kHz, $g\in[-18, 18]$, and $q\in[.1, 10]$. All values are sampled uniformly at random to produce $16,384$ train, $2048$ validation and $2048$ test signals, all $5$ seconds long. At train, validation, and test time we truncate the system response to $512$ taps.

\subsection{Results \& Discussion}
\begin{figure}[!t]
    \centering
    \includegraphics[width=.97\linewidth]{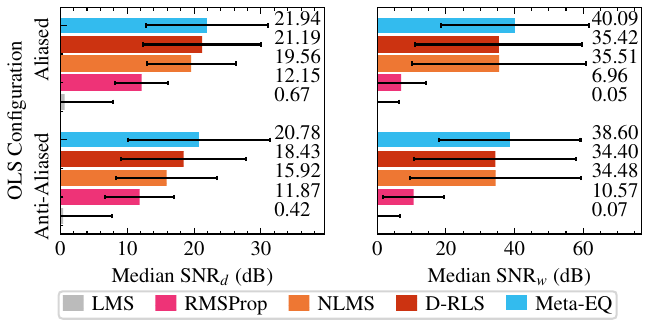}
    \caption{Equalization results for signal~($\operatorname{SNR}_d$) and system~($\operatorname{SNR}_w$) SNR. Meta-EQ performance is the least impacted by constraints.}
    \label{fig:eqbar}
\end{figure}
\begin{figure}[!t]
    \centering
    \includegraphics[width=.97\linewidth]{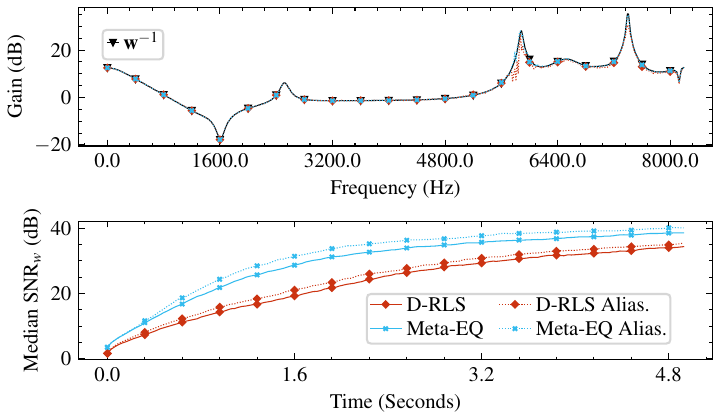}
    \caption{Comparison of true and estimated systems over time. The Meta-EQ system rapidly fits to the correct inverse model. The top plot shows an example system and the bottom shows $\operatorname{SNR}_w$ over time across the test set.}
    \label{fig:eqtime}
\end{figure}
We find our approach~(blue, solid) outperforms LMS, RMSProp, NLMS, and D-RLS for our equalization task by a noticeable margin as shown in~\fref{fig:eqbar} and further verify with a qualitative analysis plot in~\fref{fig:eqtime}.

\subsubsection{Constrained vs Unconstrained}
For the unconstrained case, our method outperforms D-RLS in $\operatorname{SNR}_d$ by $.75\,$dB and by $4.67\,$dB in $\operatorname{SNR}_w$. When we look at the constrained case, the performance for all models is degraded. Interestingly, however, our performance is proportionally degraded the least. We hypothesize that our approach learns to perform updates which are aware of the constraint.

\subsubsection{Temporal Performance Analysis}
We display final system and convergence results in \fref{fig:eqtime}. Our Meta-EQ model finds better solutions more rapidly than D-RLS. D-RLS diverged $\approx300$ times but Meta-EQ never did. 

\subsubsection{Computational Complexity}
Our learned AF has a single CPU core RTF of $\approx0.24$, and $32\,$ms latency. Our optimizer network alone has $\approx14K$ complex-valued parameters and single CPU core RTF of $\approx0.19$.
\section{Dereverberation Ablation}\label{sec:dereverb}

\subsection{Problem Formulation}
\begin{figure}[!t]
    \centering
    \includegraphics[width=.97\linewidth]{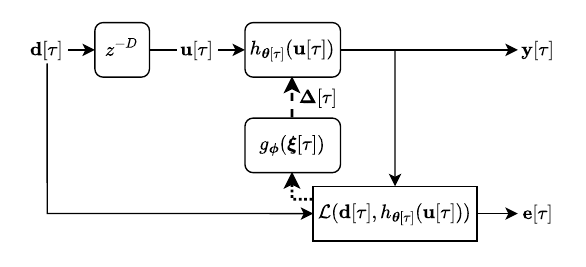}
    \vspace{-5mm}
    \caption{Prediction block diagram. A buffer of past inputs are used to estimate a future, unknown signal. The delay, $z^{-D}$ signifies a delay by $D$ frames.}
    \label{fig:wpe}
\end{figure}

For our fourth task, we train a Meta-AF to perform dereverberation via multi-channel linear prediction~(MCLP) or the weighted prediction error~(WPE) formulation, as is commonly used for speech-to-text pre-processing.
The WPE formulation is based on the idea of being able to predict the reverberant part of a signal from a linear combination of past samples, most commonly in the frequency-domain~\cite{yoshioka2009adaptive, nakatani2010speech} and shown as a block diagram in~\fref{fig:wpe}. Using our method, we use a multi-channel linear frequency-domain filter $h_{\btheta}$ and adapt the filter weights $\btheta$ using a learned AF $g_{\bphi}$ to minimize the normalized ISE AF loss below.

Assuming an array of $M$ microphones, we estimate a dereverberated signal with a linear model via
\begin{equation}
\hat{s}_{\freq\mic}[\frame] = d_{\freq\mic}[\frame] - \w_{\freq}[\frame]^\H \u_{\freq}[\frame]
\label{eq:wpe_error}
\end{equation}
where $\hat{s}_{\freq\mic}[\frame] \in \mathbb{C}$ is the current dereverberated signal estimate at frequency $\freq$ and channel $\mic$, $d_{\freq\mic}[\frame] \in \mathbb{C}$ is the input microphone signal, $\mathbf{w}_{\freq}[\frame] \in \mathbb{C}^{BM}$ is a per frequency filter with $B$ time frames and $M$ channels flattened into a vector, and $\u_{\freq}[\frame] \in \mathbb{C}^{BM}$ is a running flattened buffer of $\d_{\freq}[\frame - D]$.

We then minimize a per channel and frequency loss
\begin{equation}
    \L(\hat{s}_{\freq\mic}[\frame], \mathbf{\lambda}_{\freq}[\frame])=\frac{\|\hat{s}_{\freq\mic}[\frame]\|^2}{\mathbf{\lambda}^2_{\freq}[\frame]},
\end{equation}
\begin{equation}
    \mathbf{\lambda}^2_{\freq}[\frame] = \frac{1}{M(B+D)} \sum_{\mathrm{n}=\frame - B - D}^\frame \d_{\freq}[\mathrm{n}]^\H \d_{\freq}[\mathrm{n}],
\end{equation}
where $\lambda^2_{\freq}[\frame]$ is a running average estimate of the signal power and $\d_{\freq}[\frame] \in \mathbb{C}^{M}$.
We use this formation within our framework to perform online multi-channel dereverberation or Meta-WPE and focus on dereverberating a single output channel.

\subsection{Experimental Design}
We compare our Meta-WPE to frame-online NARA-WPE~\cite{drude2018nara}, a BD-RLS based AF which uses the WPE formulation. We ablate the filter size and inputs across: $M=1,4,8$ microphones~(optimizee size and input modification). We seek to test if our method can scale from single- to multi- channel tasks without modification. We train a new $g_\bphi$ for each $M$~(no tuning).
We measure performance with two metrics, segmental speech-to-reverberation ratio~($\operatorname{SRR}$)~\cite{wung2020robust} and $\operatorname{STOI}$. $\operatorname{SRR}$ is a signal level metric and measures how much energy was removed from the signal. It is computed as,
\begin{equation}
    \operatorname{SRR}(\d_\freq[\frame], \hat{\s}_{\freq}[\frame]) = 10 \cdot \log_{10} \left(\frac{\|\hat{\s}_{\freq}[\frame]\|^2}{\|\d_{\freq}[\frame]-\hat{\s}_{\freq}[\frame]\|^2} \right),
    \label{eq:wpe_snr}
\end{equation}
where smaller values indicate more removed energy and better performance. $\operatorname{STOI}$ is computed between the dereverberated signal estimate and the ground truth anechoic signal. We train $g_\bphi$ via~\aref{alg:inner_outer} on two GPUs, which took at most $24\,$h. We use an OLA filter with a Hann window size of $N=512$ samples and a hop of $R=256$ samples on $16\,$kHz audio. We fix the buffer size $B=5$ taps and the delay to $D=2$ frames.

We use the simulated REVERB challenge dataset~\cite{kinoshita2016summary}. The REVERB challenge contains echoic speech mixed with noise at $20\,$dB in small~(T60 = $.25\,$sec), medium~(T60 = $.5\,$sec) and large~(T60 = $.7\,$sec) rooms at near and far distances. The array is circular with a diameter of $20\,$cm. Background noises are generally stationary. The dataset has $7861$ training files, $1484$ validation files, and $2176$ test files.

\subsection{Results \& Discussion}
We find our approach~(blue, solid) outperforms NARA-WPE in $\operatorname{SRR}$ across all filter configurations, but is worse in $\operatorname{STOI}$ as shown in~\fref{fig:wpeall}. We discuss this below.

\subsubsection{Overall and Temporal Performance}
\begin{figure}[!t]
    \centering
    \includegraphics[width=.97\linewidth]{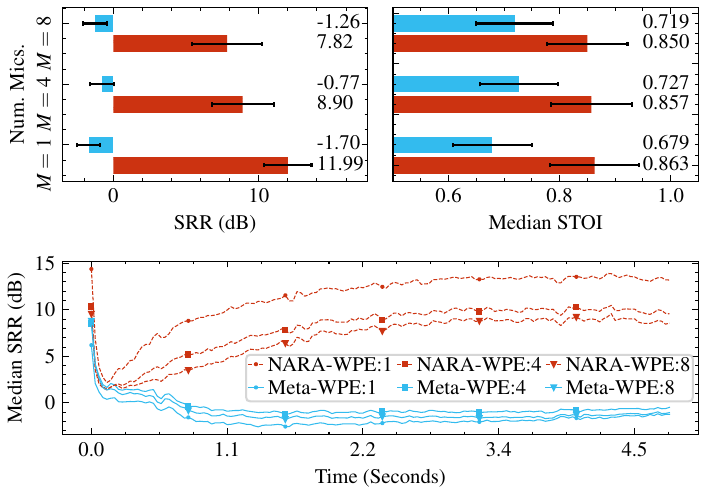}
    \caption{Dereverberation performance in terms of $\operatorname{SRR}$. Meta-WPE excels in $\operatorname{SRR}$, a metric which measures energy removed. However, in $\operatorname{STOI}$, Meta-WPE scores worse.}
    \label{fig:wpeall}
\end{figure}

As shown in~\fref{fig:wpeall}, Meta-WPE~(blue, solid) scores strongly on $\operatorname{SRR}$, where our single-channel Meta-WPE model scores better than $4$ and $8$ channel NARA-WPE~(red, dotted) models. However, as shown by $\operatorname{STOI}$, the perceptual quality is poor. While Meta-WPE is solving the prediction more rapidly, as shown by segmental $\operatorname{SRR}$, it is not doing so in a perceptual pleasing manner. Previous studies~\cite{jukic2016adaptive, wung2020robust} have also encountered this phenomenon, and propose a variety of regularization tools to align the instantaneous optimization objective with perceptually pleasing processing. We re-ran these experiments with a buffer of size $B=10$ as well as with larger and smaller optimizer network capacities and found this trend was consistent. As a result, we conclude our approach is very effectively improving the online optimization of the target loss, but the instantaneous loss itself needs to be changed to better align with perception.

\subsubsection{Computational Complexity}
The $4$ channel learned AF has a single CPU core RTF of $\approx0.47$, and $16\,$ms latency. Our Meta-WPE optimizer network alone has $\approx17K$ complex-valued parameters and single CPU core RTF of $\approx0.38$.
\section{Beamforming Ablation}\label{sec:beamforming}

\subsection{Problem Formulation}
\begin{figure}[!t]
    \centering
    \includegraphics[width=.97\linewidth]{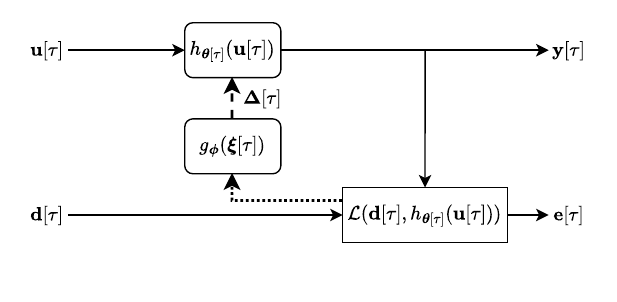}
    \vspace{-5mm}
    \caption{Informed interference cancellation block diagram. An auxiliary signal is used as input to an adaptive filter which is fit to an alternate signal.}
    \label{fig:gsc}
\end{figure}

For our fifth and final task, we train a Meta-AF for interference cancellation using the minimum variance distortionless response~(MVDR) beamformer. The MVDR beamformer can be implemented as an AF using the generalized sidelobe canceller~(GSC) \cite{gannot2017consolidated} formulation and is commonly used for far-field voice communication and speech-to-text pre-processing. We depict a version of this problem setup in~\fref{fig:gsc} and use a linear frequency-domain filter $h_{\btheta}$. We use the mixture $\d[\frame]$ as the target, informed input $\u[\frame]$, and adapt the filter weights $\btheta[\frame]$ using our learned Meta-GSC AF $g_{\bphi}$ and ISE AF loss.

Assuming an array of $M$ microphones, we have the time-domain signal model for mic $\mic$ via,
\begin{equation}
    \underline{u}_{\mic}[\time] = \underline{r}_{\mic}[\time] \ast \underline{s}[\time] + \underline{n}_{\mic}[\time]
\end{equation}
where $\underline{u}_{\mic}[\time] \in \mathbb{R}$ is the input signal, $\underline{n}_{\mic}[\time] \in \mathbb{R}$ is the noise signal, $\underline{s}[\time] \in \mathbb{R}$ is the target signal, and $\underline{r}_{\mic}[\time] \in \mathbb{R}$ is the impulse response from the source to mic $\mic$. In the time-frequency domain with a sufficiently long window, this can be reformulated as
\begin{equation}
    u_{\freq\mic}[\frame] =  r_{\freq\mic}[\frame] s_{\freq}[\frame] + n_{\freq\mic}[\frame].
\end{equation}

The GSC beamformer also assumes access to a steering vector, $\mathbf{v}_{\freq}$. While estimating the steering vector is well studied~\cite{gannot2017consolidated}, it remains non-trivial for real-world applications. For our case, we assume access to a clean speech recording $\mathbf{s}_{\freq}[\frame]$ and first compute 
\begin{equation}
    \mathbf{\Phi}^{ss}_{\freq}[\frame] = \gamma \mathbf{\Phi}^{ss}_{\freq}[\frame - 1] + (1 - \gamma)(\mathbf{s}_{\freq}[\frame]\mathbf{s}_{\freq}[\frame]^\H + \lambda \I_\mathrm{M})
\end{equation}
where $\mathbf{\Phi}^{ss}_{\freq}[\frame] \in \mathbb{C}^{M \times M}$ is a time-varying estimate of the target signal spatial covariance matrix, $\gamma$ is a forgetting factor, and $\lambda$ is a regularization parameter. We then estimate the steering vector by computing the normalized first principal component of the target source covariance matrix,
\begin{align}
    \tilde{\mathbf{v}}_{\freq}[\frame] &= \mathcal{P}(\mathbf{\Phi}^{ss}_{\freq}[\frame])\\
    \mathbf{v}_\freq[\frame] &= \tilde{\mathbf{v}}_\freq[\frame] /\tilde{v}_{\freq0}[\frame]
\end{align}
where $\mathcal{P}(\cdot)$ extracts the principal component and $\mathbf{v}_{\freq}[\frame] \in \mathbb{C}^M$ is the final steering vector. 
We then use the steering vector to estimate a blocking matrix $\mathbf{B}_{\freq}[\frame]$. The blocking matrix is orthogonal to the steering vector and can be constructed as
\begin{equation}
    \mathbf{B}_{\freq}[\frame] = 
    \begin{bmatrix}
    -\frac{[v_{\freq 1}[\frame], \cdots, v_{\freq M}[\frame]]^\H}{v_{\freq0}[\frame]^\H}\\
    \I_{M - 1 \times M - 1}
    \end{bmatrix} \in \mathbb{C}^{M \times M - 1}.
\end{equation}
The distortionless constraint is then satisfied by applying the GSC beamformer as
\begin{equation}
    \hat{s}_{\freq}[\frame] = (\mathbf{v}_{\freq}[\frame] - \mathbf{B}_{\freq}[\frame] \w_{\freq}[\frame])^\H \u_{\freq}[\frame]
\end{equation}
where $\w_{\freq}[\frame]\in \mathbb{C}^{M - 1}$ is the adaptive filter weight, and the desired response for the loss is $d_{\freq}[\frame]=\mathbf{v}_{\freq}[\frame]^\H \u_{\freq}[\frame]$. 

Our objective is to learn an optimizer $g_\bphi$ that minimizes the AF ISE loss using this GSC filter implementation. By doing so, we learn an online, adaptive beamformer that listens in one direction and suppresses interferers from all others.

\subsection{Experimental Design}
We compare our Meta-GSC beamformer to LMS, RMSProp, NLMS, and BD-RLS beamformers, in scenes with either diffuse or directional noise sources. We seek to test if our method can learn to process scenes with different spatial characteristics without modification. We train a single $g_\bphi$ and tune all baselines on a single dataset of all scene types.
We measure performance using scale-invariant source-to-distortion ratio~($\operatorname{SI-SDR}$)~\cite{le2019sdr} and $\operatorname{STOI}$. $\operatorname{SI-SDR}$ is computed as,
\begin{align}
    \mathbf{a} &= (\underline{\hat{\mathbf{s}}}^\T \underline{\mathbf{s}})/\|\underline{\mathbf{s}}\| \\
    \operatorname{SI-SDR}(\underline{\mathbf{s}}, \underline{\hat{\mathbf{s}}}) &= 10 \cdot \log_{10}(\|\mathbf{a}\underline{\mathbf{s}}\|^2/\|\mathbf{a}\underline{\mathbf{s}} -\underline{\hat{\mathbf{s}}}\|^2), 
    \label{eq:gsc_snr}
\end{align}
where larger values indicate better performance. $\operatorname{STOI}$ is computed between the output and desired speech signal. We also compute the BSS eval metrics, source-to-distortion ratio~($\operatorname{SDR}$), source-to-interference ratio~($\operatorname{SIR}$), and source-to-artifact ratio~($\operatorname{SAR}$)~\cite{vincent2006performance}. We train $g_\bphi$ via~\aref{alg:inner_outer} on one GPU, which took $\approx24\,$h. We use an OLA filter with a Hann window size of $N=1024$ samples and a hop of $R=512$ samples on $6$ channel $16\,$kHz audio.

We use the CHIME-3 challenge proposed in \cite{barker2015third, barker2017third}. This dataset contains scenes with simulated speech and relatively diffuse noise sources in a multi-channel environment. The array is rectangular and has six microphones spaced around the edge of a smart-tablet. There are $7,138$ training files, $1,640$ validation files, and $1,320$ test files. When running this dataset with directional sources, we mix spatialized speech from one mixture with the spatialized speech from a random other mixture. We do not mix speech across folds.

\subsection{Results and Discussion}
\begin{figure}[!t]
    \centering
    \includegraphics[width=.97\linewidth]{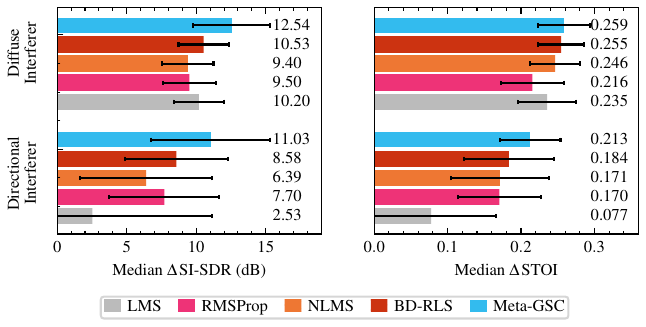}
    \caption{Performance comparison across interferers. The directional noise is the most challenging and diffuse is the easiest.}
    \label{fig:gscbar}
\end{figure}

\begin{figure}[!t]
    \centering
    \includegraphics[width=.97\linewidth]{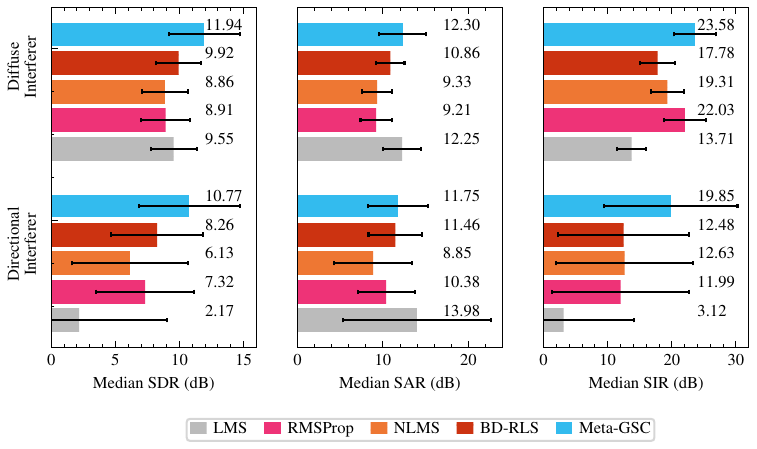}
    \caption{BSS eval comparison across interferers. Meta-GSC provides more suppression with less distortion and artifacts.}
    \label{fig:gscbssbar}
\end{figure}

\begin{figure}[!t]
    \centering
    \includegraphics[width=.97\linewidth]{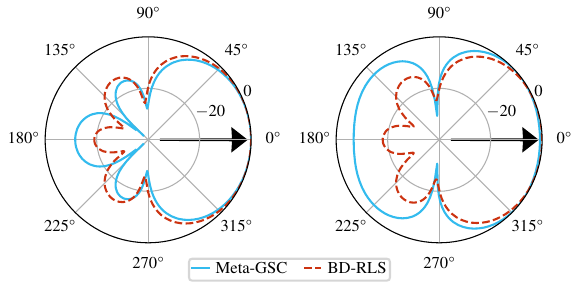}
    \caption{Spatial response plots at $\approx1\,$KHz for a directional interferer at $\approx 1$ sec.~(left) and $\approx 2$ sec.~(right).}
    \label{fig:gscbeam}
\end{figure}

We find that Meta-GSC outperforms LMS, RMSProp, NLMS, and BD-RLS in median performance metrics as shown in~\fref{fig:gscbar} and \fref{fig:gscbssbar} and in a qualitative analysis in~\fref{fig:gscbeam}.

\subsubsection{Diffuse Interferers}
The diffuse scenario tests the ability to suppress omnidirectional noise in a perceptually pleasant fashion. We show these comparisons in the ``Diffuse Interferer'' rows of~\fref{fig:gscbar} and~\fref{fig:gscbssbar}. The median input $\operatorname{STOI}$ was $0.675$ and the median input $\operatorname{SI-SDR}$ was $-0.67$. Meta-GSC~(blue, solid) outperforms BD-RLS~(red, dotted) in $\operatorname{SI-SDR}$ performance with Meta-GSC scoring $12.54\,$dB improvement and BD-RLS scoring $10.53\,$dB improvement. In $\operatorname{STOI}$, Meta-GSC outperforms BD-RLS by $0.004$. The BSS Eval metrics show that Meta-GSC provides $5.8\,$dB more interferer suppression~($\operatorname{SIR}$) while simultaneously introducing $1.4\,$dB fewer artifacts~($\operatorname{SAR}$) and $2.02\,$dB less distortion~($\operatorname{SDR}$) than BD-RLS. Typically, enhancement algorithms trade improved interference suppression for additional artifacts. However, the meta-training scheme produces an optimizer which simultaneously improves both.

\subsubsection{Directional Interferers}
The directional scenario tests the ability to suppress sources from one particular direction -- typically achieved by steering nulls in the beam pattern. We show these comparisons in the ``Directional Interferer'' rows of~\fref{fig:gscbar} and~\fref{fig:gscbssbar}. The median input $\operatorname{STOI}$ was $0.734$ and the median input $\operatorname{SI-SDR}$ was $-0.45$. Meta-GSC scores $11.03\,$dB on $\operatorname{SI-SDR}$ improvement whereas BD-RLS scores $8.58\,$dB. $\operatorname{STOI}$ performance trends similarly with Meta-GSC outperforming BD-RLS by $0.029$. The BSS-Eval metrics show a similar trend with Meta-GSC providing $7.37\,$dB more $\operatorname{SIR}$ while simultaneously introducing $.29\,$dB fewer artifacts~($\operatorname{SAR}$) and $2.51\,$dB less distortion~($\operatorname{SDR}$) than BD-RLS. We hypothesize Meta-GSC steers sharper nulls and learns an automatic VAD-like controller.

\subsubsection{Beampattern Comparison}
We compute beam plots for Meta-GSC and BD-RLS at $\approx 1$ sec. and $\approx 2$ sec. in a scene with a directional interferer. As expected, the models share the same look direction. However, our Meta-GSC method appears to steer more aggressive nulls as shown in \fref{fig:gscbeam}

\subsubsection{Computational Complexity}
Our Meta-GSC method has a single CPU core RTF of $\approx0.54$, and $32\,$ms latency. The optimizer network alone has $\approx14K$ complex-valued parameters and single CPU core RTF of $\approx0.25$. 
\section{Discussion, Future Work, and Conclusion}

\subsection{Discussion}
When we review the cumulative results of our approach, we note several interesting observations. First, the performance of our meta-learned AFs is strong and compares favorably to conventional optimizers across all tasks. Second, the performance difference between our meta-learned AFs and conventional AFs is larger for tasks that are traditionally more difficult to model by humans including AEC double-talk, AEC path changes, and directional interference cancellation. Third, we found that we could use a single configuration of our method for all five tasks, which significantly reduced our development time. This suggests that our learned AFs are a viable replacement of human-derived AFs for a variety of audio signal processing tasks and are most valuable for complex AF tasks that typically require more human engineering effort.

When we reflect on how our learned optimizers are able to achieve this success, we note two core reasons. First, and most obvious, our meta-learned AFs are \emph{data-driven} and trained on a particular class of signals~(e.g. speech, directional noise, etc). Thus, Meta-AF is limited by the capacity of our optimizer network and training data and not signal modeling skill. Second, by framing AF development as a meta-learning problem, we effectively distill knowledge of our meta loss into the AF loss and corresponding learned update rules, thus enabling us to learn AFs which optimize objectives that would be very difficult~(e.g. our frame accumulated) or even impossible to develop manually~(e.g. supervised losses, STOI, SI-SDR, etc).

\subsection{Future Work}
The field of meta-learning and meta-learned optimizers is young and has a bright future for signal processing applications. 
Future directions of research include improving training methods, non-linear optimizee filtering, optimizer architecture, and the optimizer/meta loss. 
Particularly promising avenues for future work including identifying better optimization losses and better filter representations for Meta-AF style optimization. 
Overall, we are optimistic that our Meta-AF approach can benefit from both adaptive filtering advances as well as deep learning progress and will be an exciting research topic.

\subsection{Conclusion}
We present a general framework called Meta-AF to automatically develop adaptive filter update rules from data using meta-learning. Our proposed approach offers several benefits including the first general-purposes method of learning AF update rules directly from data and a self-supervised training algorithm that does not require any supervised labeled training data. To demonstrate the power of our framework, we test it on all four canonical adaptive filtering architectures and five unique tasks including system identification, acoustic echo cancellation, equalization, dereverberation, and GSC-based beamforming -- all using a single configuration trained on different datasets. In all cases, we were able to train high performing AFs, which outperformed conventional optimizers as well as certain state-of-the-art methods. We are excited about the future of deep learning combined with adaptive filters and hope our complete code release will stimulate further research and rapid progress.

\section{Acknowledgments}
This work was generously supported by Adobe Research. We would also like to thank the anonymous reviewers for their feedback and comments, which greatly improved our manuscript.

\bibliographystyle{IEEEtran}
\bibliography{IEEEabrv,references}

\end{document}